\documentstyle[prd,preprint,tighten,aps,eqsecnum,
newlfont,epsfig]{revtex}
\begin{document}
\draft
\preprint{
\begin{tabular}{r}
DFTT 5/97
\\
hep-ph/9703415
\end{tabular}
}
\title{INELASTIC $\mathbf{\nu}$ AND
$\mathbf{\overline\nu}$ SCATTERING ON NUCLEI
AND ``STRANGENESS'' OF THE NUCLEON }
\author{
W.M. Alberico$^{\mathrm{a}}$,
M.B. Barbaro$^{\mathrm{a}}$,
S.M. Bilenky$^{\mathrm{b}}$,
J.A. Caballero$^{\mathrm{c},\mathrm{d}}$,
C. Giunti$^{\mathrm{a}}$,
C. Maieron$^{\mathrm{a}}$,
E. Moya de Guerra$^{\mathrm{c}}$ and
J.M. Ud\'{\i}as$^{\mathrm{c},\mathrm{f}}$
}
\address{
\begin{tabular}{c}
$^{\mathrm{a}}$INFN, Sezione di Torino
and Dipartimento di Fisica Teorica, Universit\`a di Torino,
\\
Via P. Giuria 1, 10125 Torino, Italy
\\
$^{\mathrm{b}}$Joint Institute for Nuclear Research, Dubna, Russia
\\
and Technion, Physics Department, 320000 Haifa, Israel
\\
$^{\mathrm{c}}$Instituto de Estructura de la Materia, CSIC, 
\\
Serrano 123, E-28006 Madrid, Spain
\\
$^{\mathrm{d}}$Permanent address: Dpto. de F\'{\i}sica At\'omica, Molecular 
y Nuclear, \\
Universidad de Sevilla, 
Apdo. 1065, E-41080 Sevilla, Spain
\\
$^{\mathrm{f}}$Present address: Dpto. de F\'{\i}sica At\'omica, Molecular y
Nuclear,\\ Fac. de CC. F\'{\i}sicas, Univ. Complutense de Madrid,\\ 
Ciudad Universitaria,
E-28040 Madrid, Spain
\end{tabular}
}
\date{\today}
\maketitle
\begin{abstract}

Possibilities to extract information on the strange form factors
of the nucleon from neutrino (antineutrino) inelastic scattering
on nuclei, in an energy range from 200 MeV to 1 GeV and more, 
are investigated
in detail. All calculations are performed within two relativistic 
independent particle models (Fermi gas and shell model); the final state 
interactions of the 
ejected nucleon are taken into account through relativistic optical model 
potentials. We have shown that the values of the cross sections 
significantly  depend on the nuclear model (especially in the lower 
energy range). However the NC/CC neutrino-antineutrino asymmetry in a
medium--high energy range shows a rather small dependence on the
model and allows to disentangle different values of the parameters
that characterize the strange form factors.
We have calculated also the ratio of the cross sections for inelastic 
NC scattering of neutrinos on nuclei, with the emission of a proton and
of a neutron. Our calculations show that this 
ratio depends rather weakly on the nuclear model and confirm previous 
conclusions on the rather strong dependence of this ratio upon the axial 
strange form factors.

\end{abstract}



\section{Introduction}
\label{INTRO}

The elastic and inelastic NC scattering of neutrinos (and antineutrinos)
on nucleons and nuclei can be an important tool to 
determine the structure of the hadronic weak neutral current.
In the present paper we will consider in detail what kind of information 
about the matrix elements of the axial and vector {\textit{strange}}
currents can be obtained from the investigation of these neutrino 
processes.

The one--nucleon matrix element of the axial
strange current,
\begin{displaymath}
<p|{\bar s}\gamma^\alpha \gamma_5 s|p> 
= 2Ms^\alpha g_A^s,
\end{displaymath}
($p$ is the nucleon momentum, $s^\alpha$ the spin vector, $M$ is the
nucleon mass  and $g_A^s$ is the strange axial constant) 
has received new attention (see \cite{Kaplan,Garv,noi})
after the measurements of the polarized structure function $g_1$ of the 
proton performed by the EMC \cite{EMC} collaboration and by the latest 
experiments done at CERN \cite{CERN} and SLAC\cite{SLAC}.
According to the theoretical analysis of these data the value 
$g_A^s= - 0.10\pm 0.03$ has been set\cite{Ellis}. Apart from 
experimental uncertainties, however, this
value is affected by several assumptions, like the small $x$ behaviour
of the polarized structure function of the proton(see for example
ref.\cite{ANSELMINO}) and the assumption of exact SU(3) 
symmetry\cite{SU3}.

The strange vector current, instead, has been somewhat investigated
in the context of parity violating electron scattering, where 
polarized electron beams are employed to disentangle the tiny 
electromagnetic--weak interference cross section. The existing 
measurements\cite{PVEX} do not allow to fix up the
so-called strange magnetic moment of the nucleon, leaving 
uncertainties even on its sign. 

These facts point to the importance of exploiting other methods for the 
determination of the matrix elements of the strange vector and axial 
currents. 

It has been pointed out in a preceding work\cite{noi} 
that measurements of the asymmetry:
\begin{equation}
{\cal{A}}_{N}(Q^2)
=
{\displaystyle\left({\displaystyle\mathrm{d} \sigma
\over\displaystyle\mathrm{d} Q^2}
\right)_{\nu N}^{\mathrm{NC}}-
\left({\displaystyle\mathrm{d} \sigma
\over\displaystyle\mathrm{d} Q^2}
\right)_{\overline\nu N}^{\mathrm{NC}}
\over\displaystyle
\left({\displaystyle\mathrm{d} \sigma
\over\displaystyle\mathrm{d} Q^2}
\right)_{\nu n}^{\mathrm{CC}}-
\left({\displaystyle\mathrm{d} \sigma
\over\displaystyle\mathrm{d} Q^2}
\right)_{\overline\nu p}^{\mathrm{CC}}}\;,
\label{asym}
\end{equation}
could allow an unambiguous determination of the presence of the 
magnetic and/or axial strange form factors of the nucleon $N$. The 
numerator of eq.(\ref{asym}) contains the difference 
between the elastic $\nu({\overline\nu})-N$ neutral current (NC) 
scattering cross sections, while in the denominator the 
difference of the cross sections of the charged current (CC)
processes $\nu_\mu(\overline\nu_\mu)+n(p)\to \mu^-(\mu^+)+p(n)$ is considered.
As it is shown in ref.\cite{noi},
using the standard model expressions for the nucleonic neutral and
charged weak currents (the former including strange currents as well) 
the expression of the asymmetry reads:
\begin{equation}
{\cal{A}}_{p(n)} =
{\displaystyle 1\over\displaystyle 4 \left| V_{ud} \right|^2}
\left(\pm 1 -{\displaystyle F^s_A \over\displaystyle F_A}\right)
\left(\pm 1 - 2 \sin^2\theta_W
{\displaystyle G_M^{p(n)} \over\displaystyle G_M^3} -
{\displaystyle G_M^s \over\displaystyle 2 \, G_M^3} \right)\;.
\label{asym2}
\end{equation}
where $G_M^{p(n)}(Q^2)$ is the magnetic form factor of the proton
(neutron), $G_M^3(Q^2)=(G_M^p-G_M^n)/2$ is the isovector nucleon magnetic
form factor, $F_A(Q^2)$ the CC axial form factor,
$V_{ud}$ is the element of the CKM mixing matrix and $\theta_W$ is 
the Weinberg angle.  In addition to these quantities,
which are relatively well known, the strange axial ($F_A^s$) and 
magnetic ($G_M^s$) form factors enter directly into (\ref{asym2}) 
and could be measured. In the above $-Q^2=q^2=q_0^2 -{\vec q}\,^2$ 
is the four--momentum transfer square.

Several present (and future [see  for example   ref.\cite{LAMPF}]) 
neutrino experiments  employ complex nuclei as a target.
Thus it is important to analyze the scattering cross sections for
inelastic $\nu({\overline\nu})$--nucleus processes. 

In this work we consider the following reactions:
\begin{equation}
\nu_\mu({\overline\nu_\mu}) + A \longrightarrow
\nu_\mu({\overline\nu_\mu}) +N + (A-1)
\qquad\qquad (NC)
\label{NCpro}
\end{equation}
\begin{equation}
\nu_\mu({\overline\nu_\mu}) + A \longrightarrow
\mu^-({\mu^+}) +p(n) + (A-1)\, ,
\qquad\qquad (CC)
\label{CCpro}
\end{equation}
where $A$ represents a nucleus with mass number $A$.
We perform a thorough analysis of the influence of various nuclear 
effects on the relative cross sections: the main task is to investigate
the possibility of extracting relevant information on the 
strange form factors. 

The theoretical estimates of the cross sections for the processes 
(\ref{NCpro}) and (\ref{CCpro}) are 
obviously affected by the nuclear model employed 
for the description of the nucleonic dynamics: since the effect of 
strange form factors is believed to be at most of the order of 
$10\div 15\%$, the uncertainty stemming from the specific nuclear
model employed must be constrained within a few percent, otherwise
the analysis of strangeness in nuclei becomes hopeless.

  An additional complication of neutrino experiments
  concerns the poor knowledge of the kinematical variables at the
  lepton vertex. For NC processes the final neutrino cannot be
  detected at all, whereas for the CC ones the final charged lepton can be
  detected and its energy and momentum could be in principle measured.
  However, in both cases the energy momentum balance cannot be precisely
  determined at the lepton vertex because of the lack of monochromatic
  neutrino beams.

  In these experiments, the energy-momentum of the ejected nucleon
   can be measured but, as the initial nucleon is bound in the target
   nucleus and the ejected nucleon interacts with the residual nucleus
   (the so called final state interactions, FSI), the energy-momentum
   balance occurring at the weak interaction vertex is not 
  unambiguously determined.

Here  we have considered the inelastic neutrino (antineutrino)--
nucleus cross section at intermediate/large energy transfers for both 
NC and CC processes. We compare the results obtained within two typical 
nuclear models, the relativistic Fermi gas (RFG) and a relativistic 
shell model (RSM), which have been widely tested in the past for, e.g.,
inelastic electron--nucleus scattering. These models can be viewed 
as two rather extreme descriptions of the nuclear structure: 
while the RFG is a very schematic model that just takes into account
the average kinetic energy of the nucleons in the nuclear medium, 
the RSM accounts for very detailed single-particle properties.
The differences in the estimated cross sections provided by the two
models can be assumed to be a fair indication of the theoretical 
uncertainty related to the nuclear model itself.

Neutrino--nucleus scattering has been considered in previous works
by Horowitz and collaborators\cite{Hor} both for neutral current 
reactions within the RFG and for charged current reactions using a
relativistic meson--nucleon model with Random--Phase--Approximation
(RPA) corrections and momentum dependent self--consistent mean field.
\cite{HorPie}. According to their conclusions RFG seems to be adequate
at relatively high momentum transfers, in agreement with
Singh and Oset\cite{Oset}, who found (non--relativistic) RPA 
corrections to be large only at low momentum transfers.
The RSM used here has been successfully tested against precise data
on coincidence elastic electron-nucleus scattering \cite{eep1,eep2}.

The main point we want to state here is that, even when the nuclear
model effects are sizable in the evaluation of separate cross 
sections, the information on strange form factors which can be 
extracted from ratios of cross sections (like the nuclear analogous 
of the asymmetry (\ref{asym})~) are 
weakly affected by the different nuclear dynamics of the models.

The explicit description of the models employed for the calculation
of the $\nu(\overline\nu)$--nucleus cross sections is introduced in
Section II, while in Section III we discuss the numerical results
together with the implications for the possibility of measuring 
the strange form factors of the
nucleon. The influence of Coulomb corrections on the charged current
cross sections and of the interaction of the ejected nucleon with
the residual nucleus is thoroughly examined and found to be not 
negligible, even for the ratios of cross sections considered here.

\section{Formalism}

For the description of the nuclear structure we employ here two
independent particle models: the Fermi gas and the Shell Model,
both of them in a relativistic  version, which occurs to be more 
appropriate when the involved energy transfers are of the order 
of several hundreds MeV.
In both cases it is assumed that the incoming neutrino (or antineutrino) 
interacts with a single nucleon, the remaining $A-1$ being spectators.

Let us start by fixing up the kinematics of the process, which is 
illustrated in Fig.1: the scattering plane $(x,z)$ is determined by the
initial (${\vec{k}}$) and final (${\vec{k}}\,'$) lepton momenta, the initial nucleus 
being at rest, and contains the momentum transfer ${\vec{q}}={\vec{k}}-{\vec{k}}\,'$.
In the Impulse Approximation (IA) the intermediate boson with momentum 
$q$ is absorbed by a single nucleon with momentum $p$ inside the
nucleus, which is then scattered to a final state with momentum $p_N$
(possibly after strong interactions with the residual nucleus):
$\vec{ p}_N$ forms an angle $\gamma$ with $\vec{ q}$, while $\phi_N$ 
is the angle between the scattering plane and the one containing $q$
and $p_N$. 

The exclusive cross sections we are interested in are generated, in lowest
order of the electroweak interaction, by the Feynman
amplitude associated with the diagram of Fig.2, where the hadronic final 
state is identified by the four-momenta of the ejected nucleon ($p_N$)
and of the daughter, A-1 nucleus ($p_{A-1}$).

We consider the initial nucleus in its ground state $\Psi_A$ 
(at rest in the laboratory frame), while the final nuclear system will
be described by the product of the knocked out nucleon wavefunction
$\psi_N(p_N, s_N)$ and the residual nucleus state, $\Psi_{A-1}$, both
of them being chosen within suitable model wavefunctions, which will
be discussed below.

The nuclear current operator is the sum of single nucleon (one--body)
currents.
For the initial and final nuclear wavefunctions an independent 
particle model (IPM) is employed; then the exact nuclear current matrix 
elements can be formally written by using an {\it effective} current 
operator (see  for example Ref. \cite{Chin92})
\begin{equation}
J_A^\mu = 
<\Psi_{A-1}^{IPM}\psi_N(p_N,s_N)|{\hat J}^\mu_{eff}|\Psi_A^{IPM}>
\label{Jeff}
\end{equation}
where all the complexities inherent to the use of {\it exact} 
wave functions have been incorporated in the unknown effective 
current operator, that in general should be a rather complicated 
many--body operator.
The matrix elements of the current are evaluated in the Impulse
Approximation (IA), where the effective current operator is substituted 
by the {\it free} one--body nuclear current operator
\begin{equation}
{\hat J}^\mu_A= \sum_{k=1}^A {\hat J}^\mu_k\, ,
\label{oneb}
\end{equation}
${\hat J}_k^\mu$ being either the neutral or the charged single nucleon
weak current operator; it is assumed to be on shell, as for the 
interaction in free space. This might be a rather crude approximation 
depending upon the kinematical conditions of the reaction under 
investigation. For the fairly large neutrino energies (and energy 
transfers) we are interested here, the IA is expected to be a reliable
approximation\cite{Fru84}.

We will now consider in more details the two independent particle 
models employed here, namely the relativistic Fermi gas and a 
relativistic shell model: they entail a quite different description
of the nuclear structure and should allow a serious test of the 
influence of the nuclear model on the quantities under investigation.

\subsection{ Relativistic Fermi Gas (RFG)}

In this paragraph we present the relevant formalism used to express the NC 
and CC inelastic scattering cross sections for the processes (\ref{NCpro})
and (\ref{CCpro}) within a relativistic Fermi gas model for the nuclear 
target. We remind that a relativistic description of the 
single nucleon states (and currents) has proved to be of some relevance
when the energy-momentum transferred to the nuclear system exceeds 
about 0.5~GeV.\cite{Bill}

Within the framework of the RFG we shall restrict ourselves to the
Plane Wave Impulse Approximation (PWIA), which does not take into account
the interaction between the knocked out nucleon and the residual nucleus, 
as it is illustrated in Fig.3. In this case three-momentum 
conservation in the laboratory system (where the initial nucleus is
at rest) implies $\vec{ p}_{A-1}=-\vec{ p}$, $\vec{ p}$ 
being the momentum
of the struck nucleon before the interaction with the leptonic current.
In the naive FG, nucleons inside the Fermi sea are on the mass--shell, 
with $p_0=\sqrt{\vec{ p}^2+M^2}$. However it is possible, without major
modifications of the approach, to account for an average, constant 
binding energy of the nucleon $-\epsilon_B$, by replacing $p_0\longrightarrow
p_0-\epsilon_B$.

The outgoing nucleon, instead, is obviously assumed to be on shell,
with energy $E_N=\sqrt{\vec{ p}^2_N+M^2} \equiv T_N + M$, $T_N$ being
its kinetic energy and $M$ the nucleon mass. We notice that in 
PWIA the relation $q=p_N-p$ also holds. 

The phase--space of the final states for the neutrino--nucleus 
scattering process is defined by the three--momenta of the outgoing
lepton, nucleon and daughter nucleus: the latter however is not
detected and thus one has to integrate the differential cross section
stemming from the amplitude represented in Fig.3 over $\vec{ p}_{A-1}$ or, 
according to the above considerations, over the struck nucleon
momentum $\vec{ p}$, whose range is constrained within the occupied
levels in the Fermi sphere ($|\vec{ p}|\le p_F$, $p_F$ being the Fermi
momentum).

We thus write the $\nu$--nucleus  differential cross 
section with respect to the energy and angles of the ejected nucleon 
as follows
\begin{eqnarray}
\left(\frac{d^2\sigma}{d E_N d\Omega_N}\right)_{\nu(\overline\nu) A}&=&
\frac{G_F^2}{(2\pi)^2} \frac{V}{(2\pi)^3}
\frac{|{\vec{p}}_N|}{4 k_0 }
\int \frac{d^3 k'}{k_0'}\frac{d^3 p}{p_0}
\delta\left(k_0-k_0'+p_0-E_N\right)
\nonumber\\
&\times& \delta^{(3)}\left({\vec{k}}-{\vec{k}}'+{\vec{p}}-{\vec{p}}_N\right)
\theta(p_F-|{\vec{p}}\,|)\theta(|{\vec{p}}_N|-p_F)
\left(L^{\mu\nu} \pm L_5^{\mu\nu}\right)
w_{\mu\nu}^{s.n.}
\label{sigmaA}
\end{eqnarray}
where $V$ is the nuclear volume\begin{footnote}{Within the Fermi 
gas model the volume of the system can be
re--expressed, via the relation $Z/V=N/V=p^3_F/3\pi^2$ (we consider here 
only $N=Z$ nuclei), in terms of the number of protons ($Z$) or neutrons
($N$) which enter into play in the specific process; at the same time 
the nucleonic form factors in the hadronic tensor will be 
specified as the ones of the corresponding nucleon (proton or neutron, 
respectively).}
\end{footnote}, 
$L^{\mu\nu}$ and $L^{\mu\nu}_5$ are the symmetric and 
antisymmetric parts, respectively, of the leptonic tensor,
\begin{eqnarray}
L^{\mu\nu} &=&
k^\mu {k'}^\nu + {k'}^\mu k^\nu - g^{\mu\nu}k\cdot k'\\
L^{\mu\nu}_5 &=&
i\epsilon^{\mu\nu\rho\sigma}k_\rho {k'}_\sigma,
\end{eqnarray}
and the plus (minus) sign refer to neutrino (antineutrino) scattering.
Finally, $w_{\mu\nu}^{s.n.}$ is the single nucleon hadronic tensor: 
\begin{equation}
w_{\mu\nu}^{s.n.} =
\sum_{s,s_N}<p_N,s_N| {\hat J}_\mu |p,s>
<p,s| {\hat J}_\nu^{\dagger}|p_N,s_N>.
\label{Whadron}
\end{equation}
In the above ${\hat J}_\mu$ is the weak nucleonic current,
with matrix elements
\begin{eqnarray}
&&<p_N,s_N|{\hat J}_\mu|p,s> =
{\overline U}_{s_N}(p_N) \Gamma_\mu U_s(p)
\label{current}\\
\qquad &&\equiv {\overline U}_{s_N}(p_N) \left[
\gamma_\mu F_V(Q^2) + \frac{i}{2M}\sigma_{\mu\alpha}q^{\alpha} F_M(Q^2)
+ \gamma_\mu\gamma_5 F_A(Q^2)- q_\mu\gamma_5 F_P(Q^2)\right] U_s(p)\, ,
\nonumber
\end{eqnarray}
where the vector ($F_V$), magnetic ($F_M$) and axial ($F_A$) nucleonic
form factors have to be specified, according whether one needs to
consider neutral or charged processes and $Q^2=-q^2$, being $q=p_N-p$.
The pseudoscalar component ($F_P$) concerns only charged
currents; in any case it does not contribute to {\it differences} of
neutrino and antineutrino cross--sections, as the ones we are interested
in for constructing an asymmetry like (\ref{asym}).

By inserting (\ref{current}) into (\ref{Whadron}) one gets the general
structure ($X_\mu=p_\mu-(p\cdot q) q_\mu/q^2$):
\begin{equation}
w_{\mu\nu}^{s.n.}= 8 M^2\left\{
-W_1\left(g_{\mu\nu}-\frac{q_\mu q_\nu}{q^2}\right)
+ \frac{W_2}{M^2} X_\mu X_\nu
+ \frac{W_3}{M^2} i\epsilon_{\mu\nu\alpha\beta} p^{\alpha} q^{\beta}
+ \frac{W_4}{M^2} q_\mu q_\nu \right\}
\label{whadron2}
\end{equation}
with:
\begin{equation}
\begin{array}{l}
W_1 = -\displaystyle{\frac{q^2}{4M^2}}
\left[ \left(F_V + F_M\right)^2 + F_A^2 \right] + F_A^2 \\
W_2 = F_V^2 -\displaystyle{\frac{q^2}{4M^2}} F_M^2 + F_A^2 \\
W_3 = F_A \left( F_V + F_M \right) \\
W_4 = - \displaystyle{\frac{M^2}{q^2}} F_A^2 +
\displaystyle{\frac{p\cdot q}{2}}F_P^2 +M F_P F_A,\\
\end{array}
\label{WWW}
\end{equation}
where the on--shell condition ($p^2=M^2$) for the struck nucleon 
inside the Fermi sea has been exploited. 

Then the $\nu({\overline\nu})$--nucleus cross section is expressed, in the 
RFG model, as follows:
\begin{eqnarray}
\left(\frac{d^2\sigma}{d E_N d\Omega_N}\right)_{\nu({\overline\nu}) A}&=&
\frac{G_F^2}{(2\pi)^2} \frac{V}{(2\pi)^3}
\frac{|{\vec{p}}_N|}{4 k_0 }\theta(|{\vec{p}}_N|-p_F)
\int \frac{d^3 k'}{k_0'}\frac{d^3 p}{p_0}
\delta\left(k_0-k_0'+p_0-E_N\right)
\nonumber\\
&\times& \delta^{(3)}\left({\vec{k}}-{\vec{k}}'+{\vec{p}}-{\vec{p}}_N\right)
\theta(p_F-|{\vec{p}}\,|)
\label{sigmaA2}\\
&\times& 8M^2 \left\{
2k\cdot k' W_1 + \frac{W_2}{M^2}\left[2(k\cdot p)\,( k'\cdot p) 
- M^2( k\cdot k')\right]\right. 
\nonumber\\
&\pm &\frac{2W_3}{M^2}
\left(k'\cdot p + k\cdot p\right) k\cdot k'
\nonumber\\
&+& \left. m_l^2 \left[W_1\frac{k\cdot k'}{q^2} +\frac{W_2}{M^2}
\left(\frac{k\cdot k'}{4}- k\cdot p\right) \mp \frac{2W_3}{M^2}
k\cdot p +\frac{W_4}{M^2} k\cdot k'\right]\right\}
\nonumber
\end{eqnarray}
where the upper (lower) sign refers to $\nu ({\overline\nu})$ induced 
processes. Obviously terms proportional to the final lepton mass 
($m_l$) only come
into play for CC processes; they are derived in (\ref{sigmaA2}) 
by exploiting the condition ${k'}^2=m_l^2$ (with $m_l=0$ for NC, 
$m_l=m_\mu$ for CC).

The integration over ${\vec{k}}'$ can be carried out by using the delta
function, while ${\vec{p}}$ can be (numerically) integrated by taking 
into account the kinematical conditions of the scattering under 
investigation. We will consider a definite value for $k$: usually 
the $\nu$ beam is not monochromatic and the analysis
of real experimental data will require an integration over the
neutrino energy spectrum; $p_N$ is measured by 
detecting the outgoing nucleon. This fixes the four--vector
$$\epsilon=p_N-k$$
and the Mandelstam variable
$$u=\epsilon^2=(k-p_N)^2=M^2-2p_N\cdot k$$
Let us define $\theta_N$ as the angle between ${\vec{k}}$ and ${\vec{p}}_N$
(remind that $|{\vec{k}}|^2=k_0^2$):
$$|{\vec{\epsilon}}\,|=|{\vec{p}}_N-{\vec{k}}|=
\left\{{{\vec{p}}\,^2_N}+k_0^2 -2|{\vec{p}}_N|k_0\cos\theta_N\right\}^{1/2}$$
Further, by assuming ${\vec{\epsilon}}$ along the $z$-axis, we call
$\theta_p$ the angle between ${\vec{p}}$ and ${\vec{\epsilon}}$ 
($\cos\theta_p={\hat p}\cdot{\hat\epsilon}$) and rewrite
the remaining energy--conserving  delta function 
in terms of $\cos\theta_p$ as follows:
$$\frac{1}{k_0'}\delta\left(k_0-k_0'+p_0-E_N\right)=
\frac{1}{|{\vec{p}}\,||{\vec{\epsilon}}\,|}\delta(\cos\theta_p-y_0)$$
having set
\begin{equation}
y_0=\frac{1}{|{\vec{p}}\,||{\vec{\epsilon}}\,|}\left(\epsilon_0 p_0 
-\frac{M^2+u-m_{l}^2}{2}\right)
\label{y0}
\end{equation}
The last expression allows to perform the integral over $d\cos\theta_p$:
this already eliminates (at least partially) the explicit dependence
upon $q^2$ of the integrand, though it will be formally maintained in the
structure factors $W_i$. Again the leptonic mass appearing in
(\ref{y0}) concerns only the charged current processes in the denominator
of the asymmetry: for neutral current $\nu$--nucleus cross sections,
obviously, it has to be set equal to zero.

The cross section (\ref{sigmaA2}) can be rewritten as:
\begin{eqnarray}
\left(\frac{d^2\sigma}{d E_N d\Omega_N}\right)_{\nu({\overline\nu}) A}&=&
\frac{G_F^2}{(2\pi)^2} \frac{V}{(2\pi)^3}
\frac{2M^2|{\vec{p}}_N|}{k_0|{\vec{\epsilon}}|}\theta\left(|{\vec{p}}_N|-p_F\right)
\int_0^{p_F}\,d|{\vec{p}}\,|\frac{|{\vec{p}}\,|}{p_0} 
\nonumber\\
&\times&\int_{-1}^1 d\cos\theta_p \delta(\cos\theta_p-y_0)
\int_0^{2\pi} d\phi
\left\{{\cal I}_1 + {\cal I}_2 \pm {\cal I}_3 + {\cal I}_4\right\}\, ,
\label{sigmaA3}
\end{eqnarray}
where, after some algebra, the functions ${\cal I}_i$ are defined
as follows:
\begin{equation}
{\cal I}_1 = - W_1(q^2)\left( M^2-u - 2k\cdot p \right)
\left(1 + \frac{m_l^2}{2q^2}\right)
\label{I1}
\end{equation}
\begin{equation}
{\cal I}_2 = \frac{W_2(q^2)}{M^2}\left[
\frac{M^2}{2}\left(M^2 - u\right) -u (k\cdot p) + 
\frac{m_l^2}{4}\left( k\cdot p -\frac{M^2-u}{2}\right)\right]
\label{I2}
\end{equation}
\begin{equation}
{\cal I}_3 = \frac{2W_3(q^2)}{M^2}\left[
(k\cdot p)\left(k\cdot p -\frac{1}{2}m_l^2\right) -
\frac{1}{4}\left(M^2-u\right)\left(M^2-u+m_l^2\right) \right]
\label{I3}
\end{equation}
and
\begin{equation}
{\cal I}_4 = \frac{W_4(q^2)}{M^2}m_l^2
\left[k\cdot p -\frac{1}{2}\left(M^2-u\right)\right]
\label{I4}
\end{equation}

In the above $q^2=M^2-u+m_l^2-2 k\cdot p$ and the scalar product
$k\cdot p$ is a non--trivial function of $\cos\theta_p$ and $\phi$.
As already stated, in the previous formulas we have employed the
on--shell condition, $p_0=\sqrt{M^2+{{\vec{p}}\,}^2}$, for the nucleon inside
the Fermi sea; this condition, however, can be partly released in
order to account for an average, constant binding energy 
$\epsilon_B$\cite{Hor}. In
this case the energy $p_0$ in the $\delta$--function in (\ref{sigmaA})
must be replaced by $p_0-\epsilon_B$. As we will show in the numerical results
presented below, this ``minor'' modification makes the RFG cross 
sections (in the quasi--elastic kinematics considered here) much more 
realistic and closer to the shell model calculation.

Another correction, which refers only to the CC processes, stems from
the distortion on the wavefunction of the final (charged) lepton due
to its interaction with the Coulomb field of the (residual) nucleus.
Still remaining in Born Approximation for the main scattering process,
this Coulomb correction should be taken into account by replacing the
plane wave describing the final lepton with an ``exact'' eigenfunction
of the nuclear Coulomb field. This procedure, however, is somewhat 
complicated (see, for example, refs.\cite{eep1,Ube}): the main effects
of the Coulomb distortion can be more easily accounted for by means of
the prescription (which can be used both within the RFG and
the RSM) described below.

The main point is to replace the plane wave, $e^{i{\vec{k}}'\cdot{\vec{r}}}$, of
the outgoing lepton by:
\begin{displaymath}
\frac{|{\vec{k}}\,'_{eff}|}{|{\vec{k}}\,'|}e^{i{\vec{k}}\,'_{eff}\cdot{\vec{r}}}
\end{displaymath}
where
\begin{equation}
{\vec{k}}\,'_{eff} = {\vec{k}}\,'\left(1 \pm\frac{3}{2}\frac{Z\alpha}{R|{\vec{k}}'|}
\right).
\label{keff}
\end{equation}
In the above the plus (minus) refers to the lepton ( $\mu^-$) and the
antilepton ($\mu^+$), respectively, $Z$ is the number of protons
and $R\simeq 1.2 A^{1/3}$ is the effective charge radius of the
nucleus under investigation. This approximation has been tested within a
non--relativistic approach in ref.\cite{GiustiPac}.

The substitution ${\vec{k}}\,'\longrightarrow {\vec{k}}\,'_{eff}$ obviously 
affects the kinematics in the three--momentum conserving $\delta$--function
and the phase space of the final lepton, 
but also the energy $k_0'$ of the outgoing lepton (still to be considered
on the mass shell), once it is expressed in terms of ${\vec{k}}\,'_{eff}$.
We shall comment upon the Coulomb corrections on the CC cross--sections
in the discussion of the numerical results.

Thus far we have considered double differential cross sections: however,
to test the sensitivity to the presence of strange form factors, we will
consider single differential cross 
sections with respect to the (kinetic) energy of the knocked out 
nucleon. The latter can be obtained from our previous formulas by 
further integrating over the solid angle $\Omega_N$. Formally:
\begin{eqnarray}
\left(\frac{d\sigma}{dE_N}\right)_{\nu (\overline\nu)A}&&=
\left(\frac{d\sigma}{dT_N }\right)_{\nu(\overline\nu)A}
= \frac{G_F^2}{2\pi} \frac{V}{(2\pi)^3}
\frac{2M^2|{\vec{p}}_N|}{k_0|{\vec{\epsilon}}\,|}\theta\left(|{\vec{p}}_N|-p_F\right)
\int_{-1}^{+1} d\cos\theta_N 
\nonumber\\
&\times&\int_0^{p_F}\frac{d|{\vec{p}}\,||{\vec{p}}\,|}{p_0} 
\theta (1-|y_0|) \int_0^{2\pi} d\phi\left.
\left\{{\cal I}_1 + {\cal I}_2 + {\cal I}_3 + {\cal I}_4\right\}
\right|_{\cos\theta_p=y_0}\, ,
\label{sigmaA4}
\end{eqnarray}
Total cross sections (integrated over the final nucleon energy) will
be utilized as well. They are defined as:
\begin{equation}
\sigma_{\nu(\overline\nu) A}=\int\, dT_N 
\left(\frac{d\sigma}{dT_N }\right)_{\nu(\overline\nu)A}.
\label{total}
\end{equation}
In the above we have presented general formulas, which are valid both 
for  NC and CC processes: for a specific calculation one should take 
care  of the following remarks:
\begin{itemize}
\item
in the case of NC cross--sections the form factors appearing in 
(\ref{WWW}) will be denoted by $F^Z_i$ ($i=V,M,A$) (while $F_P$ does
not contribute).
\item
in the case of CC cross-sections the form factors will be
$F_i^{CC}$ ($i=V,M,A,P$) and the additional replacement
$G_F^2 \Longrightarrow G_F^2 |V_{ud}|^2$ is required.
\end{itemize}

\vskip 1.0truecm

\subsection{ Relativistic Shell Model }

In this approach we shall use a relativistic shell model for the
wavefunctions of the initial (target) and of the final (residual) nucleus,
while we shall assume a final scattering state for the knocked out nucleon.
Moreover the IA will be employed, as in the previous case.
Thus the nuclear current matrix element $J_A^\mu$ for the
process we are interested here, is computed  as [see also eq.(\ref{Jeff})]
\begin{equation}
J_A^\mu=<\Psi^{SM}_{A-1}(P_{A-1})\psi_N(P_N,s_N)|
\sum_{k=1}^A \hat{J}^{\mu}_k|\Psi_A^{SM}(P_A)>
\label{SMme}
\end{equation}
where the (free) current operator is built up as in (\ref{oneb}) and
$P_A (P_{A-1})$ are the four--momenta of the initial (final) nucleus,
respectively.
The final state $|\Psi^{SM}_{A-1}(P_{A-1})\psi_N(P_N,s_N)>$
is a single-channel optical-model wave function  constructed from
the product of a final state for the A-1 particles  residual nucleus
 and either a plane  wave (PWIA) or a distorted wave (DWIA) for
the outgoing ejected nucleon. The initial nuclear state can be
equivalently rewritten as $|\psi_B^{SM},\Psi_{A-1}^{SM}(P_{A-1})>$,  
denoting a (bound--state) single--particle shell model wave function 
coupled to the rest of the initial nucleus.  

After performing the angular momentum algebra involved
in the shell model description for the residual and target nuclei,
the nuclear current can be expressed in terms of spectroscopic
amplitudes $f_j(I_A,I_{A-1})$ times  single--particle
current matrix elements, $I_A (I_{A-1})$ being the angular momentum
of the target (residual) nucleus.

The required single--particle matrix elements are of the form: 
\begin{equation}
J_{\mu}({\vec{q}}\,)= \sqrt{V}\int d^3r\, e^{i{\vec{q}}\cdot{\vec{r}}}\,
{\overline\psi}_{s_N}({\vec{p}}_N,{\vec{r}}\,) 
\Gamma_{\mu} \psi^{jm}_{B,\kappa}({\vec{r}}\,) \; ,
\label{spcur}
\end{equation}
where $\psi^{jm}_{B,\kappa}, \psi_{s_N}$ are the 
wave functions for the initial bound nucleon (with quantum numbers
$j,m,\kappa$)  and for the final outgoing nucleon (with momentum ${\vec{p}}_N$), 
respectively; 
$\Gamma_{\mu}$ is the same single--particle  current operator 
for free  nucleons, which was defined in eq.(\ref{current}).

We will consider first the so--called PWIA, thus neglecting the 
interaction in the final state (FSI) between the ejected nucleon and 
the residual nucleus; the single particle matrix elements of the current
can then be computed as: 
\begin{equation}
J_{\mu}({\vec{q}}\,)=\frac{1}{\sqrt{2 E_N}} {\overline U}({\vec{p}}_N,s_N)
\int d{\vec{r}} e^{-i {\vec{p}}\cdot {\vec{r}}}\Gamma_\mu \psi_{B,\kappa}^{jm}({\vec{r}}\,)
\label{269}
\end{equation}
with ${\vec{p}}={\vec{p}}_N-{\vec{q}}$.

In this approximation the differential cross-section with respect to
the energy of the ejected nucleon can be written as \cite{irep,prcyas}
\begin{eqnarray}
\frac{d\sigma^{SM}}{dE_N}&=& 4 \pi |f_j(I_A,I_{A-1})|^2  
\int\!\! d k_0' \int\!\! d(\cos \theta) E_{A-1} 
\frac{1}{|{\vec{q}}|}\left(\frac{d\sigma}{d\Omega}\right)^{Z^0/W^\pm}
\nonumber\\
&\times&
 \left[ \omega_L {\overline W}_L+ \omega_T {\overline W}_T+ 
\omega_{TT'}{\overline W}_{TT'}
\right]
\label{rsm}
\end{eqnarray}
where the bars over the structure functions $W_i$ imply averages (sum) over 
the initial (final) nucleonic states:
\begin{displaymath}
{\overline W}_i = \frac{1}{2j+1}\sum_{m,m'} W_i\, .
\end{displaymath}
In the previous expressions ${\vec{q}}={\vec{k}}-{\vec{k}}'$ is the
3--momentum transfer and $\theta$ is the
scattering angle of the final lepton. $E_{A-1}$ is the energy of
the residual nucleus, and $(d\sigma/d\Omega)^{Z^0/W^\pm}$ are  
Mott--like cross--sections that
assume the following form for neutral and charged current processes:
\begin{eqnarray}
\left(\frac{d\sigma}{d\Omega}\right)^{Z^0} & = 
&\frac{G_F^2}{2\pi^2}\:{ k_0'}^2 \cos^2( \theta/2) \\
\left(\frac{d\sigma}{d\Omega}\right)^{W^\pm} & = 
&  4 |V_{ud}|^2 \frac{G_F^2}{2\pi^2} {|{\vec{k}}'|}^2.
\end{eqnarray}
For closed shell nuclei in the extreme shell model,
we have $|f_j(I_A,I_{A-1})|^2=2j+1$. When several shells 
contribute  we just sum the corresponding cross-section
for every shell.

The integrations in Eq.(\ref{rsm}) are performed numerically.
To compute the limits, one should keep in mind that $k_0-k_0'=\omega$,
~~$\omega_{min}=M_{A-1}+E_N-M_A$ being the minimum required energy
transfer; moreover the following kinematical constraints hold:
\begin{eqnarray}
|\vec{P}_{A-1}-{\vec{p}}_N|& \le & |{\vec{q}}| \le |\vec{P}_{A-1}+{\vec{p}}_N|  \\
|{\vec{k}}-{\vec{k}}'| & \le& |{\vec{q}}| \le |{\vec{k}} +{\vec{k}}'|
\end{eqnarray}
The remaining terms in Eq. (\ref{rsm}) are the  
 ``kinematical'' coefficients $\omega_L,\omega_T,\omega_{TT'}$ 
and the response functions $W_L,W_T,W_{TT'}$, which will be specified
below.

In the case of neutral currents, since  the mass of both the initial
and final lepton is zero (so that $k_0=|{\vec{k}}|, k_0'=|{\vec{k}}'|$)
the expressions of the kinematical coefficients read:
\begin{eqnarray}
\omega_L&=&1 \\
\omega_T&=& \tan^2 (\theta/2)- \frac{q^2}{2\vec{q\,}^2} \\
\omega_{TT'} &=& \pm 2 \tan(\theta/2)
\sqrt{\tan^2(\theta/2)-q^2/\vec{q\,}^2} 
\end{eqnarray} 
the minus (plus) sign referring to neutrino (antineutrino). 

To write  explicitly the structure functions 
we choose a coordinate system defined by the
unit vectors 
(${\hat q}$, $\hat{n}_\perp$, $\hat{n}_\parallel$), where
$\hat{n}_\perp$ is
the direction perpendicular to the nucleon scattering plane 
(i.e., the plane defined by ${\vec{q}}$ and ${\vec{p}}_N$), 
and $\hat{n}_\parallel$ is a vector in the nucleon scattering plane 
perpendicular to ${\vec{q}}$ and $\hat{n}_\perp$.
We write here the four--vector hadronic current as $J^\mu\equiv
(\rho,\vec{J})$; then, according to the above definitions:
\begin{eqnarray}
{\vec J}&=& J_q \hat{q}+J_\perp \hat{n}_\perp+ J_\parallel \hat{n}_\parallel \\
\hat{n}_\perp     & \equiv & (-\sin \phi_N,\cos \phi_N,0) \\
\hat{n}_\parallel & \equiv & (\cos \phi_N,\sin \phi_N,0) \\
\vec{p}_N & \equiv & |\vec{p}_N| (\sin \gamma \cos \phi_N,\sin 
\gamma \sin \phi_N,\cos \gamma) 
\end{eqnarray}
where $\gamma,\phi_N$ are the scattering and azimuthal  angles for the 
ejected nucleon (see also Fig.~1). 

After integrating on the unobserved final nucleon angles, the structure 
functions can be rewritten in terms of the nuclear current components
as follows: 
\begin{eqnarray}
W_L&=& |\rho|^2 +\frac{\omega^2}{\vec{q\,}^2} |J_q|^2 - 
\frac{\omega}{|{\vec{q}}\,|}2 Re\:(\rho
J_q^\dagger) \label{stl}\\
W_T&=& |J_\parallel|^2 + |J_\perp|^2 \\
W_{TT'}&=&Im\: (J_\parallel J_\perp^\dagger)
 \label{slt'}
\end{eqnarray}

For charged current reactions of type $(\nu_{l},l N)$ or 
$(\overline{\nu}_{l},{\overline l} N)$ with final lepton 
mass $m_l$, we obtain the same expression for $W_T$ and 
$W_{TT'}$, while
\begin{eqnarray}
\omega_L W_L &=&\frac{1}{4k_0|{\vec{k}}'|} \left\{ \:\left[
(k_0+k_0')^2-|\vec{q\:}|^2-m_l^2 
\right] |\rho|^2 \right. 
\nonumber \\
& & \;\;\;\; +\:\left[
\frac{(k_0^2-|{\vec{k}}'|^2)^2}{|\vec{q\:}|^2} - 
\omega^2+ m_l^2 \right] |J_q|^2 
\nonumber \\
& & \left. \;\;\;\;\;\;\;\; -\:\left[
\frac{2(k_0+k_0')(k_0^2-|{\vec{k}}'|^2)}{|{\vec{q}}\:|} 
- 2\omega |{\vec{q}}\:| \right] Re\:(\rho J_q^\dagger) \right\} 
\label{slCC}\\
\omega_T  &=&   
\frac{k_0 |{\vec{k}}'| \sin^2\theta}{2|{\vec{k}}\:|^2}-\frac{1}{2}
\left(\frac{-k_0'}{|{\vec{k}}'|}+\cos\theta \right)  
\label{stCC}  \\
\omega_{TT'} &=& \pm \frac{1}{|{\vec{k}}\:|}
\left(\frac{k_0 k_0'}{|{\vec{k}}'|}+|{\vec{k}}'|- 
\bigl(k_0 + k_0'\bigr)\cos \theta \right)  
\label{sttCC}
\end{eqnarray}

In the expressions for the $TT'$  contribution, the upper and 
lower signs correspond to neutrino and anti--neutrino scattering,
respectively. 
 We have written the expressions for neutral and charged current
simultaneously, but we have to keep in mind (as it was pointed out
in the analogous situation for the RFG) that the  current
operator  to be used in Eq.(\ref{269}) 
will include different form--factors for neutral and
charged current.

 The bound nucleon wave functions are computed in a relativistic
 framework, as solutions of  a Dirac equation with scalar and 
 vector potentials.
 We use the wave functions obtained with the  TIMORA code \cite{timora}.
 This code implements a self--consistent Hartree procedure
 with Mean Field solutions of a linear
 Lagrangian  including  nucleons and
 scalar ($\sigma$)  vector--isoscalar ($\omega$) and vector--isovector 
($\rho$) mesons. The free parameters of the lagrangian (the nucleon--meson 
 coupling constants
 for the $\sigma,\omega$ and $\rho$ 
 and the mass of the scalar particle) are adjusted to reproduce 
 nuclear matter properties and the {\em rms} radius of $^{40}$Ca 
 \cite{advances}.
 Several other lagrangians including a non--linear
 self--coupling of the scalar meson, with  parameters adjusted to
 the binding energies and {\em rms} radii of magic nuclei 
 are also being currently used \cite{Reinhardt,Ring}. The 
 bound state wave functions obtained with these alternative models are
 not very different from the ones used here \cite{mitesis}.
 The solutions of the linear lagrangian used in this work well 
 reproduce the observed cross-sections of the $(e,e'p)$
 reactions in several nuclei without further
 adjustment \cite{eep1,eep2}.

 We have also verified that the results for the cross--sections of
 this work are almost insensitive to the choice of the lagrangian
 used, providing that the same single--particle binding 
 energies are used in the different models. As these ones determine
 the threshold of the cross--section for every shell, we have used 
 the experimentally measured values of the binding energies.

Within this relativistic framework the bound state wave function 
for the initial nucleon, $\psi^{jm}_{B,\kappa}$, 
is a four--spinor with well defined 
angular momentum quantum numbers $j,\, m$ and $\kappa$ corresponding 
to the shell under consideration. We use four--spinors of the form
\begin{equation}
\psi^{jm}_{B,\kappa} ({{\vec{r}}}\,)=\left(\begin{array}
{@{\hspace{0pt}}c@{\hspace{0pt}}} 
g_{\kappa}(r) 
\phi^{jm}_\kappa(\hat{r}) \\ 
 i f_{\kappa}(r) \phi^{jm}_{-\kappa}(\hat{r})\end{array}\right)
\label{psimuk}
\end{equation}
that are eigenstates of total angular momentum with eigenvalue
$j=|\kappa|-1/2$,
\begin{equation}
\phi^{jm}_{\kappa}(\hat{r}) = \sum_{m_\ell,s} 
< \ell \ m_\ell \ {{\scriptstyle{\frac{1}{2}}}} \ s | \ j \ m >  Y_{\ell m_\ell}(\hat{r}) 
\chi_{s}
\label{fimuk}
\end{equation}
with $\ell=\kappa$ for $\kappa > 0 $, $\ell=-\kappa -1$ for $\kappa < 0$.
$f_\kappa$ and $g_\kappa$ are the solutions of the usual radial 
equations \cite{Rose}. The normalization we use is $\int_V
\psi^{jm\dagger}_\kappa(\vec{r}\,) \psi^{jm}_\kappa 
(\vec{r}\,)\, d\vec{r}=1$. 

In the framework of the RSM we will consider two
different situations for the ejected nucleon: in the first one, which
compares with the RFG calculation, no interaction is taken into account
between it and the residual nucleus (PWIA). Then 
the wave function for the outgoing nucleon is also a four--component 
spinor,  obtained as a partial wave expansion in configuration 
space of a plane wave, {\em i.e.}, a solution of the {\em free}
Dirac equation. 

In the second case the FSI between the outgoing nucleon and the 
residual nucleus is accounted for by an appropriate Relativistic 
Optical Model potential (hereafter referred to as ROP), which is 
embodied in the Dirac Equation for the ejected nucleon (DWIA):
\begin{equation}
\left[ i{\vec\alpha}\cdot\vec\nabla -\beta(M+U_S) +E 
-U_V -U_C\right]\psi({\vec{r}})=0.
\label{Diraceq}
\end{equation}
The scalar ($U_S$), vector ($U_V$) and Coulomb ($U_C$) components
of the potential can be derived within the same meson--exchange
relativistic model which is employed for the description of the 
bound nuclear states; more often, however, one utilizes 
phenomenological potentials, which are fitted to the elastic 
nucleon--nucleus scattering. In particular the real part of the
(complex) optical potential is related with the elastic rescattering 
of the ejected nucleon, while the imaginary part accounts for the 
absorption of it into unobserved channels (or its 
re--absorption by the residual nucleus). 
The vector and scalar part of the ROP we employed here have the form:
\begin{eqnarray}
U_V(r,E)&=& V_0(E)f_0(r,E) +i\left[ W_0(E)g_0(r,E) + 
W_{0SP}(E)h_0(r,E)\right]
\label{vectorpot}\\
U_S(r,E)&=& V_s(E)f_s(r,E) +i\left[ W_s(E)g_s(r,E) + 
W_{sSP}(E)h_s(r,E)\right]
\label{scalarpot}
\end{eqnarray}
where $f_i$ and $g_i$ ($i=0,s$) are symmetrized Woods--Saxon functions, 
while the $h_i$ are derivatives of Woods--Saxon functions. The strengths 
$V_i, W_i$ (as well as the radii of the Woods--Saxon distributions) have
an explicit energy dependence. This ROP corresponds to one of the 
energy--dependent 
parameterization of Cooper {\em et
al.} \cite{CHCM} of  Dirac optical potentials fitted to  elastic proton 
scattering data in
an extensive range of proton energies and mass number nuclei. The 
potential used in this work is the A--independent single nucleus
parameterization for $^{12}C$ presented in ref. \cite{CHCM}. The
results obtained with  the other choices for the
optical potential  contained in ref.\cite{CHCM}  are very similar 
to the ones presented here.

Once inserted the above ROP into the Dirac equation, the corresponding 
solutions read:
\begin{equation}
\psi_N({{\vec{r}}\,})=4 \pi \sqrt{\frac{E_N+M}{2E_N}}
{\displaystyle \sum_{\kappa,m_\ell,m}}
 e^{-i\delta^{*}_{\kappa}}\,
i^{\ell} < \ell \ m_\ell \ {{\scriptstyle{\frac{1}{2}}}} \ s_N | j \ m >
Y_{\ell m_\ell}^{*}(\hat{p}_N)\psi_{\kappa}^{jm}({{\vec{r}}}\,) \; ,
\label{expan}
\end{equation}
where $\psi_{\kappa}^{jm}({{\vec{r}}}\,)$ are four--spinors of the same 
form as that in Eq. (\ref{psimuk}), except that now the radial 
functions $f_\kappa$, $g_\kappa$ are complex because of the complex 
optical potential. It should also be mentioned that since the wave 
function (\ref{expan}) corresponds to an outgoing nucleon, 
we use the complex conjugates of the radial functions and phase 
shifts (the latter with the minus sign). 

To obtain the cross--section we integrate analytically over  all 
possible (unobserved) angles for the outgoing nucleon.
The remaining integrations (on the kinematical variables
of the unobserved  final lepton) are performed numerically. 
Notice that this procedure differs from the one adopted in the RFG,
where the kinematical variables of the final lepton is integrated first.

\vskip 1.0truecm
\subsection{Nucleonic form factors}

In a previous work\cite{noi} concerning the asymmetry (\ref{asym}) 
in the context of elastic $\nu(\overline\nu)$--nucleon scattering, the
influence of different parameterizations of the electromagnetic form
factors has been carefully analyzed, being one of the sources of 
uncertainty in disentangling the presence of strange form factors at
intermediate/large $Q^2$ values. In particular the magnetic form 
factors are the only ones entering into play 
in the differences of neutrino and antineutrino
cross sections and ref.\cite{noi} shows, in the analysis of (\ref{asym}),
a non--negligible error band due to the present experimental uncertainty
in the measurements of magnetic form factors (specifically the ones of 
the neutron). 

In the present work we intend to evaluate not only the above mentioned
asymmetry, but also the separate cross sections which enter into its
definition: thus both electric and magnetic nucleonic form factors
enter into the definition of the vector and magnetic weak NC and
CC form factors. 

We are interested in values of the form factors at relatively small 
$Q^2$ ($Q^2\le 1$~GeV$^2$): in this region the standard dipole 
parameterization of the electromagnetic form factors $G_E^p$, $G_M^p$ 
and $G_M^n$ (with a dipole mass $M_V=0.84$~GeV)  and the Galster 
parameterization\cite{Galster} for the neutron electric form factor, 
$G_E^n$, provide a fair description of the experimental data\cite{Petratos}.

For simplicity the usual dipole form has also been used 
for the axial nucleonic form factor, $F_A$, with a cutoff mass 
$M_A=1.032$~GeV. Moreover the pseudoscalar form factor $F_P^{CC}$,
entering into the CC cross sections, is taken as it is given by
PCAC and pion dominance\cite{Lew}:
\begin{equation}
F_P^{CC}(Q^2) =-\frac{2M}{m_\pi^2+Q^2} F_A^{CC}(Q^2),
\label{Fp}
\end{equation}
$m_\pi$ being the pion mass.

There remain to be considered an explicit form for the strange form
factors entering into the NC weak nucleonic current: we shall use 
here the following dipole forms:
\begin{eqnarray}
G_M^s(Q^2)&=&\frac{\mu_s}{(1+Q^2/M_V^2)^2}
\label{GMs}\\
F_A^s(Q^2)&=&\frac{g_A^s}{(1+Q^2/M_A^2)^2}
\label{FAs}
\end{eqnarray}
with typical values for $g_A^s$ and $\mu_s$, which will
be discussed later. 
Alternative $Q^2$ dependences of these form factors have been 
considered in ref.\cite{noi} and could be employed here as well,
with similar outcomes.

\vskip 1.0truecm

\section{Results and discussion}

In this Section we shall present the results of our calculations for
$\nu({\overline\nu})$--nucleus cross sections using both nuclear models 
described above; in addition to separate cross section, we will also
consider the ratio between ($\nu,p$) and ($\nu,n$) NC cross sections
\begin{footnote}
{In this Section the labels of the cross sections explicitly indicate 
the ejected, final particle (proton, neutron or generic nucleon).}
\end{footnote}:
\begin{equation}
R_{p/n}={\displaystyle
{\left({\displaystyle\mathrm{d} \sigma
\over\displaystyle{\mathrm{d}} T_N}\right)_{(\nu, p)}^{\mathrm{NC}} }
\over\displaystyle
{\left({\displaystyle\mathrm{d} \sigma
\over\displaystyle{\mathrm{d}} T_N}\right)_{(\nu, n)}^{\mathrm{NC}} }}
\label{ratiopn}
\end{equation}
Moreover we shall evaluate a quantity analogous to eq.(\ref{asym})
for the case of inelastic cross sections: 
\begin{equation}
{\cal A}_{N}(T_N) = {\displaystyle
\left({\displaystyle\mathrm{d} \sigma
\over\displaystyle{\mathrm{d}} T_N}\right)_{(\nu, N)}^{\mathrm{NC}} - 
\left({\displaystyle\mathrm{d} \sigma
\over\displaystyle{\mathrm{d}} T_N}\right)_{(\overline\nu, N)}^{\mathrm{NC}}
\over\displaystyle\left({\displaystyle\mathrm{d} \sigma
\over\displaystyle{\mathrm{d}} T_N}\right)_{(\nu,p)}^{\mathrm{CC}}
- \left({\displaystyle\mathrm{d} \sigma\over\displaystyle
{\mathrm{d}} T_N}\right)_{(\overline\nu,n)}^{\mathrm{CC}} }\;,
\label{inelasymm}
\end{equation}
$T_N$ being, as usual, the kinetic energy of the ejected nucleon
(proton or neutron). 

The shell model calculations have been 
done for the $^{12}$C nucleus, while in the Fermi Gas model we employ a 
Fermi momentum $p_F=225$~MeV/c, which is supposed to account for the
average density of $^{12}$C (smaller than the ordinary nuclear matter
density).

We start by considering a relatively ``low'' incoming neutrino energy,
$E_\nu=200$~MeV, which is a typical value for the beam of neutrinos from 
decay in flight available at LAMPF\cite{LAMPF}: Figs.~4a,b show the 
$\nu(\overline\nu)+A\to \nu(\overline\nu)+p +(A-1)$ (a) 
and $\nu_\mu(\overline\nu_\mu)+A\to \mu^-(\mu^+)+ p(n) +(A-1)$ (b) 
cross sections, eq.(\ref{sigmaA4}), evaluated with the RFG model without
($\epsilon_B=0$) and with ($\epsilon_B=-25$~MeV) binding energy for the
hole states, as well as with the RSM formalism (which accounts for
the experimental binding of the occupied states). Strange form factors
are set to be zero. One can see rather
large discrepancies between the various curves; when the average 
binding is taken into account in the RFG, the latter approaches 
the RSM cross sections, though there remain differences of the order
of 10\% or more.

A separate discussion is required by the inclusion of the appropriate
corrections on the propagation of the final particles, namely the
interaction of the ejected nucleon with the residual nucleus (FSI) 
and the Coulomb interaction of the final muon, for the CC processes.
We remind here that there is a Coulomb correction also for an ejected
proton, and this is taken into account in the Optical Potential, 
together with the effects of strong interactions.

A sizable reduction of the $\nu-A$ cross sections is induced by the
FSI of the ejected nucleon, as it can be seen in Fig.~5 for the 
$\nu+A\to \nu +p+(A-1)$ process: here the 
effect of the relativistic optical model potential turns out in a
reduction of more than 50\% as compared with the corresponding PWIA
calculation. Similar results are found when the ejected nucleon is
a neutron. Together with the ROP of eqs.~(\ref{vectorpot}) and 
(\ref{scalarpot}), we have also employed an optical potential
based on the same mesonic model which describes the initial nuclear 
bound states: it provides a slightly smaller reduction than ROP, but
gives a fair description of the main effects of the FSI; the 
comparison between the two optical potentials accounts for the
theoretical uncertainty which one can ascribe to this part of the
process. 

Concerning the Coulomb distortion on the final $\mu^\pm$ in the CC
process, we have found that, at 200~MeV incident neutrino energy,
this correction produces a increase from 5 to $19\%$ in the 
$(\nu_\mu,\mu^-)$ cross sections and an decrease from 4 to $14\%$ in the 
$({\overline\nu_\mu},\mu^+)$ ones, depending upon the outgoing 
nucleon energy. Thus the correction associated with 
the Coulomb interaction of the charged lepton is not negligible. 

It is worth noticing that at this low neutrino energy the 
evaluation of the asymmetry (\ref{inelasymm}) does not seem to be of
particular interest, for at least two reasons: i) the dependence 
upon the nuclear model employed for the calculation of the separate
cross sections remains quite large in the asymmetry, thus preventing 
the use of 
(\ref{inelasymm}) to distinguish between different strangeness contents;
ii) the energy range (in $T_N$) in which the ratio between NC and CC 
differences of 
cross sections is fairly stable (constant) is quite small ($\le 20\div 
30$~MeV). This last point is due to the muon mass, which rapidly brings
the CC cross sections down to zero, contrary to what happens in the
NC cross sections.

Then we have calculated the ratio of NC $\nu$--induced cross sections
with a proton and a neutron in the final state, 
eq.(\ref{ratiopn}). This quantity was first suggested as a probe for 
strange form factors in ref.\cite{Krew,Garvey}.
It is shown first in Fig.~6 without strange form 
factors, again for an incoming neutrino energy of 200~MeV: in this figure
we want to display the sensitivity of $R_{p/n}$ to the nuclear model
description, both in the initial and in the final states. 
These results deserve the following comments: 
 
i) Within the PWIA a small difference
(of the order of $6\div 7\%$) remains between the shell 
model calculation and 
the free Fermi gas one, which in turn is much less affected by the
average binding energy, as compared to the situation in Figs.~4a,b.
These findings are in agreement  with  the calculations of Barbaro
{\textit{et al.}}\cite{Bai}, where the RFG is  compared with predictions 
of the  so--called Hybrid Model (again an independent particle approach).

ii) The inclusion of FSI, using the ROP model, leads to a sizable increase
of $R_{p/n}$, the correction becoming larger with increasing energies; 
of course both $\sigma_{\nu,p}$ and $\sigma_{\nu,n}$ are strongly 
reduced by the FSI, but by a different amount. Indeed it might be
interesting to notice that, by artificially switching off the Coulomb 
potential in (\ref{Diraceq}), the correction of the FSI on the ratio
$R_{p/n}$ becomes much smaller, being confined within about $4\%$ 
with the exception of the largest values of $T_N$.
Of course this does not imply that the Coulomb term is the most 
important in the optical model potential, since the main effects 
are ascribed to the strong interaction; the latter, however, due to
isospin invariance, are similar on protons and neutrons and tend to cancel
in the ratio of the cross sections, while the Coulomb correction does not; 
the point we wish to stress here is the necessity to take into 
account carefully any ``Coulomb distortion'' on the ejected proton.

As a final comment on this figure, we wish to compare with the results
of Garvey {\textit {et al.}}\cite{Garvey}; they evaluate the cross sections
within a continuum random phase approximation (RPA) model, the initial 
nuclear (ground) state being a Slater determinant of Woods--Saxon single 
particle wave functions. The RPA correlations provide a microscopic 
description of the FSI, while in our case the FSI are  
embodied in the phenomenological optical 
potential, and we only consider one--nucleon emission.
Even though both ways of treating FSI are very different,
once the ratio $R_{p/n}$ is considered, the two approaches
lead to similar conclusions, stressing once more the stability of this 
quantity against differences in the nuclear models employed.

The modification of the ratio $R_{p/n}$ induced by the presence of 
strange form factors is illustrated in Fig.~7, where the shell model 
calculations both in PWIA and DWIA (evaluated with the full ROP) are 
reported. We compare the ratio obtained in the absence of strangeness
($g_A^s=\mu_s=0$)  with two cases: one with $g_A^s=-0.15$ and
$\mu_s=0$, the second with $g_A^s=-0.15$ and $\mu_s=-0.3$.
It is clearly seen that the effects of strangeness are quite sizable,
particularly the ones associated with $g_A^s$; the deviations induced
by the magnetic strange form factor are smaller and comparable
with the non--negligible correction produced by the FSI. Thus, 
a measurement of $R_{p/n}$ should allow to disentangle altogether 
a contribution
from strange form factors; this conclusion is in qualitative 
agreement with the results of Garvey {\textit {et al.}}\cite{Garvey}.

We turn now to analyze situations corresponding to higher incident
neutrino energies; in particular we have considered $E_\nu=500$~MeV
and $E_\nu=1$~GeV as typical values for the discussion of the nuclear
effects on the asymmetry (\ref{inelasymm}), which is the
main focus of this work. 

Figs.~8a,b show the separated cross sections for NC and CC processes
at $E_\nu=500$~MeV: the RSM (solid lines) is again compared (in PWIA) 
with the two ``versions'' of the RFG, with (dashed lines) and without
(dot--dashed lines) binding energy: there 
remain some discrepancies between the two approaches, but limited 
within some 7\% for the RFG with binding. 
The reduction of the cross sections induced by FSI
(again incorporated by using the ROP within the shell model approach), 
instead, remains sizable (of the order of 40\%), as it appears from the
long--dashed lines. 

For this value of $E_\nu$ it starts being of some interest to consider
the asymmetry (\ref{inelasymm}), which is illustrated in Fig.~9: here we
compare ${\cal A}_p$ evaluated a) in the RFG with $\epsilon_B=-25$~MeV
(dashed lines), b) in the RSM without FSI (solid lines) and in
DWIA, with the FSI provided by the ROP (dot-dashed lines). The 
differences between the various models turn out to be quite reduced
in the ratio defining the asymmetry, with respect to the corresponding
effects on the separated cross sections. The largest correction (within
8\%) remains the one associated to the FSI. 

The finite muon mass, which, as already noticed above, 
brings the CC cross sections down to zero
at lower $T_N$ values with respect to the NC ones, produces a rapid
increase of the asymmetry for $T_N\ge 150$~MeV, thus leaving a 
reasonable energy range ($50\le T_N\le 150$~MeV) in which the 
asymmetry has a fairly constant value. Therefore it is worth comparing
the results for the asymmetry, obtained with different estimates for
the parameters ($g_A^s,\mu_s$) of the strange form factors: Fig.~9 
shows that a measurement of the asymmetry could appreciably reveal
the existence of non--vanishing axial and/or magnetic strange
form factors. Indeed the differences in ${\cal A}_N$ associated with,
e.g., a value of $g_A^s=-0.15$ amount to about 15\%, which is outside
the ``theoretical uncertainty'' provided by the excursion in ${\cal A}_N$ 
values obtained with different nuclear models.

Similar considerations apply to the $E_\nu=1$~GeV case (and to 
higher neutrino energies): here, however, it is worth noticing that 
already for the separated cross sections the differences between RFG 
and RSM turn out to be negligible, while the reduction produced by the
FSI remains sizable. This reduction, due to the imaginary term of the
optical potential, takes into account that only 
$\simeq 50\%$ of the events
correspond to the quasielastic channel. This is in rough agreement with
a Montecarlo simulation and experimental observations \cite{Ahr}.
The FSI produces, as expected, a much smaller
effect on the asymmetry, as it is illustrated in Fig.~10: only the
calculations within the RSM (without and with FSI) are shown in the 
figure, since the
corresponding curves obtained within the RFG model would practically 
coincide with the ones displayed in the figure. 
The FSI produce a correction of the
order of few ($4\div 5$) $\%$, but for the smallest $T_N$ values,
while the effects of non--vanishing strange form factors is quite larger.
Moreover we display the results obtained by taking into account,
in addition to the FSI, also
the Coulomb distortion of the final muon in the denominator of
(\ref{inelasymm}). The corresponding correction (with respect of the PWIA)
is smaller than the one associated with FSI, both of them 
resulting in a reduction of the asymmetry. 
The global effect of FSI+Coulomb distortion does not exceed about 6\% 
(again with the exception of the smallest $T_N$ values).
In any case the effects of
the strange form factors remain well distinguished with respect to 
the nuclear medium corrections. 

On the basis of the results obtained for $E_\nu=1$~GeV, we can also 
state that for higher neutrino energies the RFG model (corrected by
FSI and Coulomb distortion) can be safely employed to compare with
the experiment: indeed, as expected, the shell structure effects 
have no influence when the energy/momentum transferred to the target
nucleus are much larger than the binding of nucleons inside. Moreover,
as stated by other authors\cite{Garvey} different types of correlations,
like for example the RPA ones, become negligible as well at large
momentum transfers.

Therefore we can conclude that for incident neutrino energies larger 
than 1~GeV the influence of the nuclear model on the neutrino 
asymmetry  ${\cal A}_N$ is rather modest and well under control: 
this is an important (although expected)
outcome, since it implies that 
the sensitivity of (\ref{inelasymm}) to the unknown components of the 
nucleonic form factors is comparable to the one discussed for the
elastic scattering\cite{noi}. The real difference between the two situations
concerns the measurable kinematic variables, in particular the fact
that $Q^2$ is no longer fixed. Thus the asymmetry measured via
inelastic scattering on nuclei (which is the most common experimental
situation) could allow, as well as the elastic processes, to disentangle
the strange components of the nucleonic weak form factors.

We considered it worthwhile to evaluate, for $E_\nu=1$~GeV, also the
ratio $R_{p/n}$; indeed this neutrino energy is close to the one 
of a previous elastic scattering experiment performed in 
Brookhaven and analyzed by Garvey \textit{et al.} in connection with
the strange axial constant\cite{Garv}. In Fig.~11 we display the ratio
(\ref{ratiopn}) evaluated with the RSM, both in PWIA (solid lines) and 
in DWIA (dot--dashed lines), utilizing the ROP. 

The differences between the two approaches turn out to be fairly 
negligible, while the effect of different values for the strangeness 
parameters $g_A^s$ and $\mu_s$ (we use here the same three choices 
employed in Fig.~7) is quite large. In contrast to the $E_\nu=200$~MeV
case, nuclear model effects do not appreciably alter this ratio, 
whereas the strange components of the nucleonic form factors produce
corrections, for example, of more than 30\% when $g_A^s$ varies from 
0 to $-0.15$, thus compelling toward a direct measurement of 
this  quantity.

Concerning the $Q^2$ dependence of the form factors, we have not 
discussed here the influence of different parameterizations of the
electromagnetic and axial form factors entering into the calculation.
Indeed we have employed in the present work the usual dipole 
parameterization both for non--strange and strange form factors, keeping
as free parameters only the strengths ($g_A^s$ and $\mu_s$) of the latter.
Due to the close similarity of the present results 
(for $E_\nu\gtrsim 1$~GeV) to the ones obtained, for the asymmetry, 
in the elastic case\cite{noi}, one should keep in
mind that the uncertainties in the electromagnetic form factors, discussed
there at length, will also affect the asymmetry defined in the inelastic
neutrino--nucleus scattering: as in the previous case, however, their
entity should not spoil the possibility of disentangling the effects of
strangeness.

Finally we consider a quantity which can be defined as ``integral asymmetry''
and is obtained from the ratio of NC and CC differences between total cross
sections [see eq.(\ref{total})~]:
\begin{equation}
{\cal{A}}^I_{N} = { { \sigma_{\nu N}^{\mathrm{NC}} - 
{\sigma}_{\overline\nu N}^{\mathrm{NC}}}
\over {\sigma_{(\nu,p)}^{\mathrm{CC}}
- \sigma_{(\overline\nu,n)}^{\mathrm{CC}} } }\;.
\label{intasymm}
\end{equation}
The integral asymmetry is displayed in Fig.~12 for $E_\nu=1$~GeV, as a
function of the axial strangeness parameter, $g_A^s$, and for two different
choices of the strange magnetic moment $\mu_s$ (0 and $-0.3$). 
The calculation of ${\cal{A}}^I_p$ has been performed both in PWIA (with 
the RFG and the RSM) and in DWIA (using the ROP): although this quantity
is obtained by integrating over the ejected nucleon energy (and thus over 
the final state), the effect of 
FSI still shows up in a reduction of about 4\% of the integral asymmetry.
However the sensitivity of ${\cal{A}}^I_p$ to the strangeness parameters is
much larger than to the nuclear model effects. The correlation between
$g_A^s$ and $\mu_s$ is clearly displayed in Fig.~12.

In conclusion, the present analysis shows that quasi--elastic 
neutrino--nucleus scattering can be conveniently utilized to look for
evidence of non--vanishing strange nucleonic form factors; we have found
that both the ratio $R_{p/n}$ and the nuclear asymmetry (\ref{inelasymm})
are appropriate quantities to be considered, as they are fairly sensitive 
to the strangeness parameters.

\vskip 2cm

\textsl{ This work was finished while one of the authors (SB) was Lady 
Davis visiting professor at the Technion. This author would like to thank
the physics Department of Technion for the hospitality.\hfill\break
We acknowledge the projects: CHRX-CT 93-0323 and 94-0562 
and also CICYT PB95-0123. }


\newpage


\begin{figure}[h]
\caption[Fig.~\ref{fig01}]{\label{fig01}
The two scattering planes  involved in  the process: the initial  and  final
neutrino momenta ($k,k'$)  are in the  $(x,z)$ plane;  the  outgoing 
nucleon  ($p_N$) in  the  inclined plane.}
\end{figure}

\begin{figure}[h]
\caption[Fig.~\ref{fig02}]{\label{fig02}
Schematic representation for  the amplitude, in Born approximation, 
of  the  neutrino--nucleus  scattering.}
\end{figure}

\begin{figure}[h]
\caption[Fig.~\ref{fig03}]{\label{fig03}
Representation of  the $\nu$--nucleus scattering in  the Impulse
Approximation.}
\end{figure}

\begin{figure}[h]
\caption[Fig.~\ref{fig04}]{\label{fig04}
The NC (a) and CC (b) differential cross sections for neutrino and
antineutrino induced processes, versus the kinetic energy of the
ejected nucleon, $T_N$, at incident $\nu({\overline\nu})$ energy 
$E_\nu=200$~MeV. The solid lines represent the  RSM calculation, 
the dashed (dot dashed) lines are the results  obtained with the  RFG
with $\epsilon_B=-25$~MeV  ($\epsilon_B=0$, respectively). Here and
in the following the differential cross sections are in
$10^{-42}$~cm$^2$~MeV$^{-1}$. }
\end{figure}

\begin{figure}[h]
\caption[Fig.~\ref{fig05}]{\label{fig05}
The NC differential cross sections for neutrino induced processes, 
versus $T_N$, at incident $\nu({\overline\nu})$ energy $E_\nu=200$~MeV. 
The solid line represents the  RSM calculation within the PWIA;  the  
other curves include the  effects of FSI: with  a meson--exchange
Optical Potential (dashed line) and  with the phenomenological ROP
(dot--dashed line). No strange form factors are included. }
\end{figure}

\begin{figure}[h]
\caption[Fig.~\ref{fig06}]{\label{fig06}
The ratio $R_{p/n}$ for NC neutrino processes, versus $T_N$, at incident  
energy $E_\nu=200$~MeV.  The solid line is  the pure RSM calculation, 
the dashed (dotted) lines are  obtained with the  RFG
with $\epsilon_B=-25$~MeV  ($\epsilon_B=0$, respectively); the
dot--dashed line  corresponds to the RSM with FSI accounted for by
the  ROP model, while the long--dashed line is obtained by switching off
the  Coulomb interaction in the ROP. No  strange form  factors  are 
included.}
\end{figure}

\begin{figure}[h]
\caption[Fig.~\ref{fig07}]{\label{fig07}
The ratio $R_{p/n}$ for NC neutrino processes, versus $T_N$, at incident
energy $E_\nu=200$~MeV.  The solid lines correspond to the RSM calculation,
the dot--dashed lines include the effect of  FSI accounted for by the
ROP model; Three different choices  of strangeness parameters  are shown,
as indicated in the figure. }
\end{figure}

\begin{figure}[h]
\caption[Fig.~\ref{fig08}]{\label{fig08}
The NC (a) and CC (b) differential cross sections for neutrino and
antineutrino induced processes, versus the kinetic energy of the
ejected nucleon, $T_N$, at incident $\nu({\overline\nu})$ energy 
$E_\nu=500$~MeV. The solid lines represent the  RSM calculation in PWIA, 
the dashed (dot dashed) lines are the results  obtained with the  RFG
with $\epsilon_B=-25$~MeV  ($\epsilon_B=0$, respectively), again in
PWIA. The long--dashed lines are the RSM calculation in DWIA, using the
ROP. }
\end{figure}

\begin{figure}[h]
\caption[Fig.~\ref{fig09}]{\label{fig09}
The  asymmetry (\ref{inelasymm}),  ${\cal A}_p$ for an ejected proton, 
versus  $T_N$,  at incident $\nu({\overline\nu})$ energy $E_\nu=500$~MeV. 
The solid lines correspond to the RSM calculation, the dashed  lines  to
the RFG with $\epsilon_B=-25$~MeV and the dot--dashed lines are in  DWIA
evaluated with ROP model. The three set of curves correspond to different 
choices  of strangeness parameters: $g_A^s=\mu_s=0$   (lower lines),
$g_A^s=-0.15$, $\mu_s=0$ (intermediate  lines) and  $g_A^s=-0.15$, 
$\mu_s=-0.3$ (upper lines). }
\end{figure}

\begin{figure}[h]
\caption[Fig.~\ref{fig10}]{\label{fig10}
The  asymmetry (\ref{inelasymm}),  ${\cal A}_p$ for an ejected proton, 
versus  $T_N$,  at incident $\nu({\overline\nu})$ energy $E_\nu=1.0$~GeV. 
The solid lines correspond to the RSM calculation, the dot--dashed lines 
are in  DWIA evaluated with ROP model, the dotted lines  represent the  RSM
corrected  by the the FSI and the Coulomb distortion of   
the muon in  the CC processes in
the denominator of  ${\cal A}_p$. The three sets of curves correspond 
again to different choices  of strangeness parameters: $g_A^s=\mu_s=0$   
(lower lines), $g_A^s=-0.15$, $\mu_s=0$ (intermediate  lines) and  
$g_A^s=-0.15$, $\mu_s=-0.3$ (upper lines). }
\end{figure}

\begin{figure}[h]
\caption[Fig.~\ref{fig11}]{\label{fig11}
The ratio $R_{p/n}$ for NC neutrino processes, versus $T_N$, at incident
energy $E_\nu=1$~GeV.  The solid lines correspond to the RSM calculation,
the dot--dashed lines include the effect of  FSI accounted for by the
ROP model; Three different choices  of strangeness parameters  are shown,
as indicated in the figure. }
\end{figure}

\begin{figure}[h]
\caption[Fig.~\ref{fig12}]{\label{fig12}
The  integral asymmetry (\ref{intasymm}),  ${\cal A}^I_p$ for an ejected 
proton, versus  $g_A^s$, at incident $\nu({\overline\nu})$ energy $E_\nu=1.0$~GeV. 
The solid lines correspond to the RSM calculation, the dashed lines to
the RFG with average binding energy $\epsilon_B=-25$~MeV, the dot--dashed
lines to the DWIA evaluated with RSM and ROP model.
The two sets of curves correspond to different choices  of the magnetic 
strangeness parameter: $\mu_s=0$ (lower lines) and  $\mu_s=-0.3$ 
(upper lines). }
\end{figure}

\newpage

\begin{minipage}[p]{\textwidth}
\begin{center}
\setlength{\unitlength}{1.0cm}
\begin{picture}(14,8)
\put(0,0){
}
\thinlines
\put(0,0){\line(1,0){10}}
\put(0,0){\line(2,3){4}}
\put(4,6){\line(1,0){4.6}}
\put(12.6,6){\line(1,0){1.4}}
\put(10,0){\line(2,3){4}}
\put(8,3){\line(1,5){1}}
\put(12,3){\line(1,5){1}}
\put(9,8){\line(1,0){4}}
\thicklines
\put(8,3){\vector(3,2){2}}
\put(9.5,4){\makebox(0,0)[rb]{$p_{N}$}}
\put(5,3){\vector(1,0){3}}
\put(6.5,2.9){\makebox(0,0)[t]{$q$}}
\put(5,3){\vector(0,1){2}}
\put(5,5){\makebox(0,0)[rb]{$k'$}}
\put(3.5,1.5){\vector(1,1){1.5}}
\put(4,1.9){\makebox(0,0)[tl]{$k$}}
\thinlines
\put(5.4,3.9){\makebox(0,0)[b]{$\theta$}}
\put(2,3){\vector(0,1){2}}
\put(2,5){\makebox(0,0)[rb]{$y$}}
\put(12,3){\vector(1,0){2}}
\put(14,3){\makebox(0,0)[lt]{$z$}}
\put(4,6){\vector(2,3){1}}
\put(5,7.5){\makebox(0,0)[rb]{$x$}}
\put(9,3.3){\makebox(0,0)[l]{$\gamma$}}
\put(12.5,4.6){\makebox(0,0)[bl]{$\phi_{N}$}}
\end{picture}
\end{center}
\end{minipage}
\begin{center}
\Large Figure~\ref{fig01}
\end{center}

\newpage

\begin{minipage}[p]{\textwidth}
\begin{center}
\setlength{\unitlength}{1.0cm}
\begin{picture}(8,6)
\thicklines
\put(0,0){\line(2,3){2}}
\put(0,0){\vector(2,3){1}}
\put(2,3){\line(-2,3){2}}
\put(2,3){\vector(-2,3){1}}
\put(5,3){\line(2,3){2}}
\put(5,3){\vector(2,3){1}}
\multiput(5.8,0)(0.2,0){3}{\line(-1,3){1}}
\multiput(8,3.8)(0,0.2){3}{\line(-3,-1){3}}
\put(3.5,2.8){\makebox(0,0)[t]{$q$}}
\put(0.8,0){\makebox(0,0)[b]{$k$}}
\put(0.8,6){\makebox(0,0)[t]{$k'$}}
\put(6.8,0){\makebox(0,0)[b]{$p_{A}$}}
\put(8,3){\makebox(0,0)[r]{$p_{A-1}$}}
\put(6.3,6){\makebox(0,0)[t]{$p_{N}$}}
%
\put(0,0){
}
\end{picture}
\end{center}
\end{minipage}
\begin{center}
\Large Figure~\ref{fig02}
\end{center}

\vspace{3cm}

\begin{minipage}[p]{\textwidth}
\begin{center}
\setlength{\unitlength}{1.0cm}
\begin{picture}(8,8)
\thicklines
\put(0,2){\line(2,3){2}}
\put(0,2){\vector(2,3){1}}
\put(2,5){\line(-2,3){2}}
\put(2,5){\vector(-2,3){1}}
\put(5,5){\line(2,3){2}}
\put(5,5){\vector(2,3){1}}
\put(5,3){\line(0,1){2}}
\put(5,3){\vector(0,1){1.3}}
\multiput(5.8,0)(0.2,0){3}{\line(-1,3){1}}
\multiput(8,3.8)(0,0.2){3}{\line(-3,-1){3}}
\put(3.5,4.8){\makebox(0,0)[t]{$q$}}
\put(0.8,2){\makebox(0,0)[b]{$k$}}
\put(0.8,8){\makebox(0,0)[t]{$k'$}}
\put(6.8,0){\makebox(0,0)[b]{$p_{A}$}}
\put(8,3){\makebox(0,0)[r]{$p_{A-1}$}}
\put(6.3,8){\makebox(0,0)[t]{$p_{N}$}}
\put(4.8,4){\makebox(0,0)[r]{$p$}}
%
\put(0,0){
}
\end{picture}
\end{center}
\end{minipage}
\begin{center}
\Large Figure~\ref{fig03}
\end{center}

\begin{minipage}[p]{\textwidth}
\begin{center}
\mbox{\epsfig{file=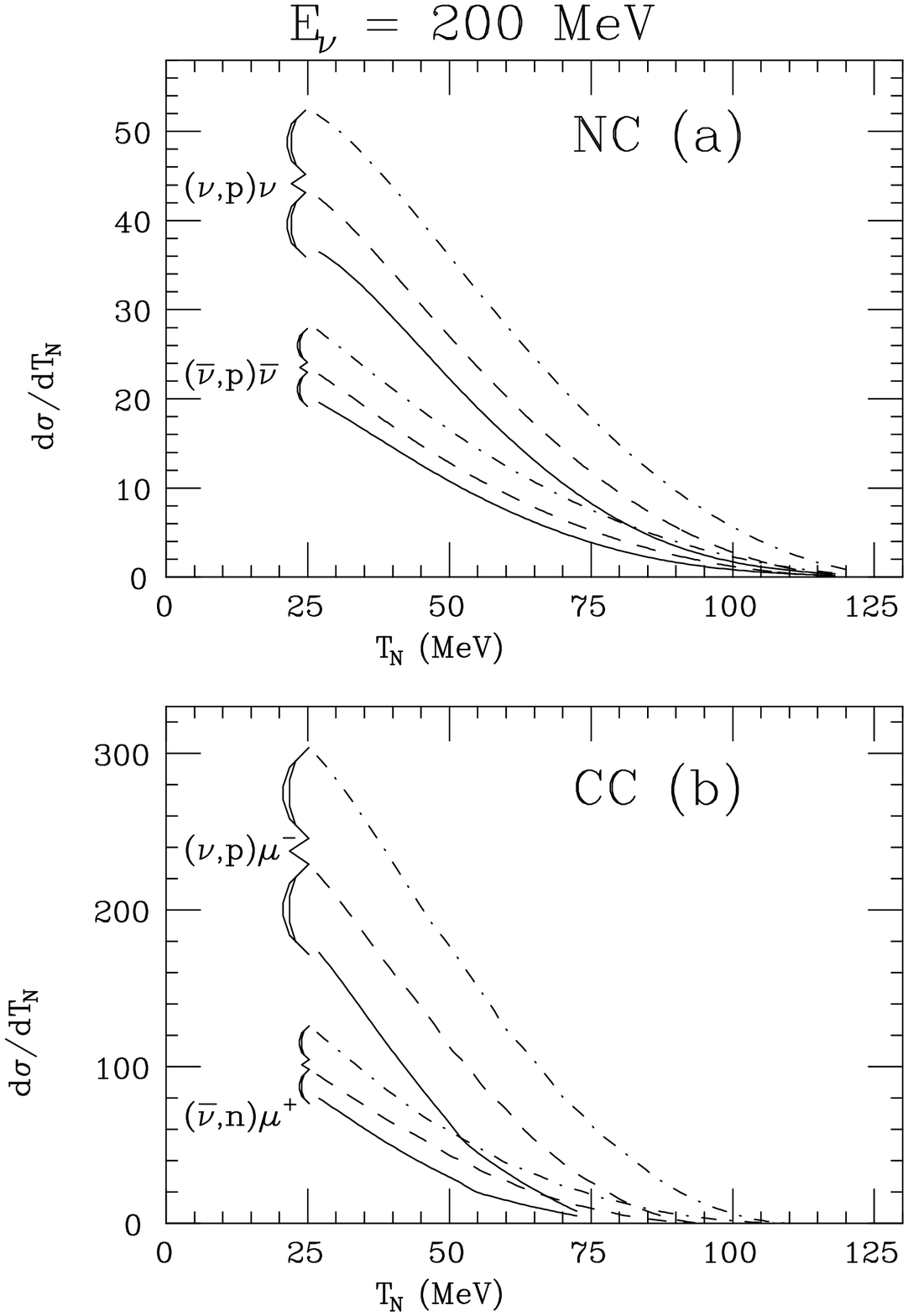,width=0.9\textwidth}}
\end{center}
\end{minipage}
\begin{center}
\Large Figure~\ref{fig04}
\end{center}

\begin{minipage}[p]{\textwidth}
\begin{center}
\mbox{\epsfig{file=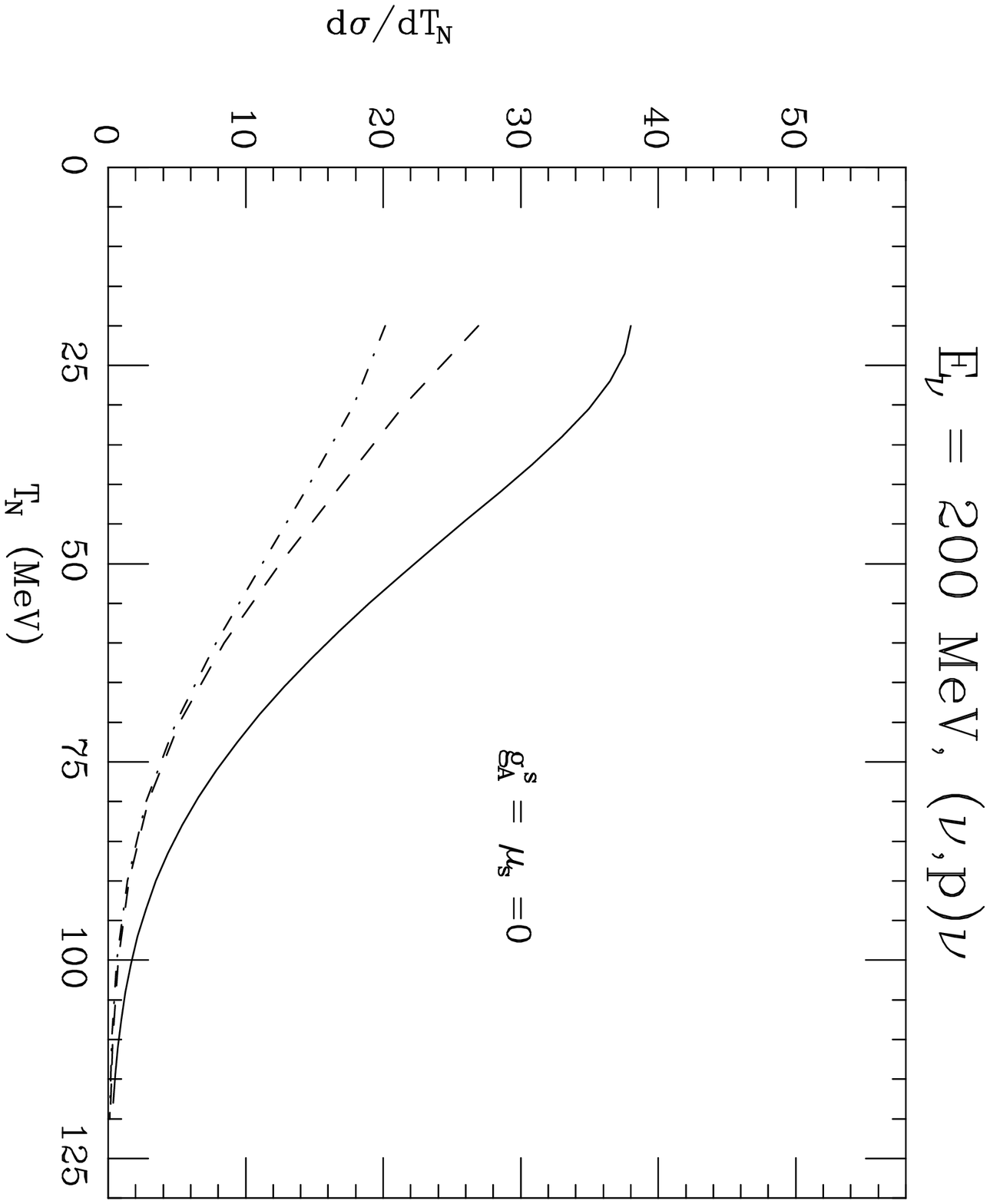,width=0.9\textwidth}}
\end{center}
\end{minipage}
\begin{center}
\Large Figure~\ref{fig05}
\end{center}

\begin{minipage}[p]{\textwidth}
\begin{center}
\mbox{\epsfig{file=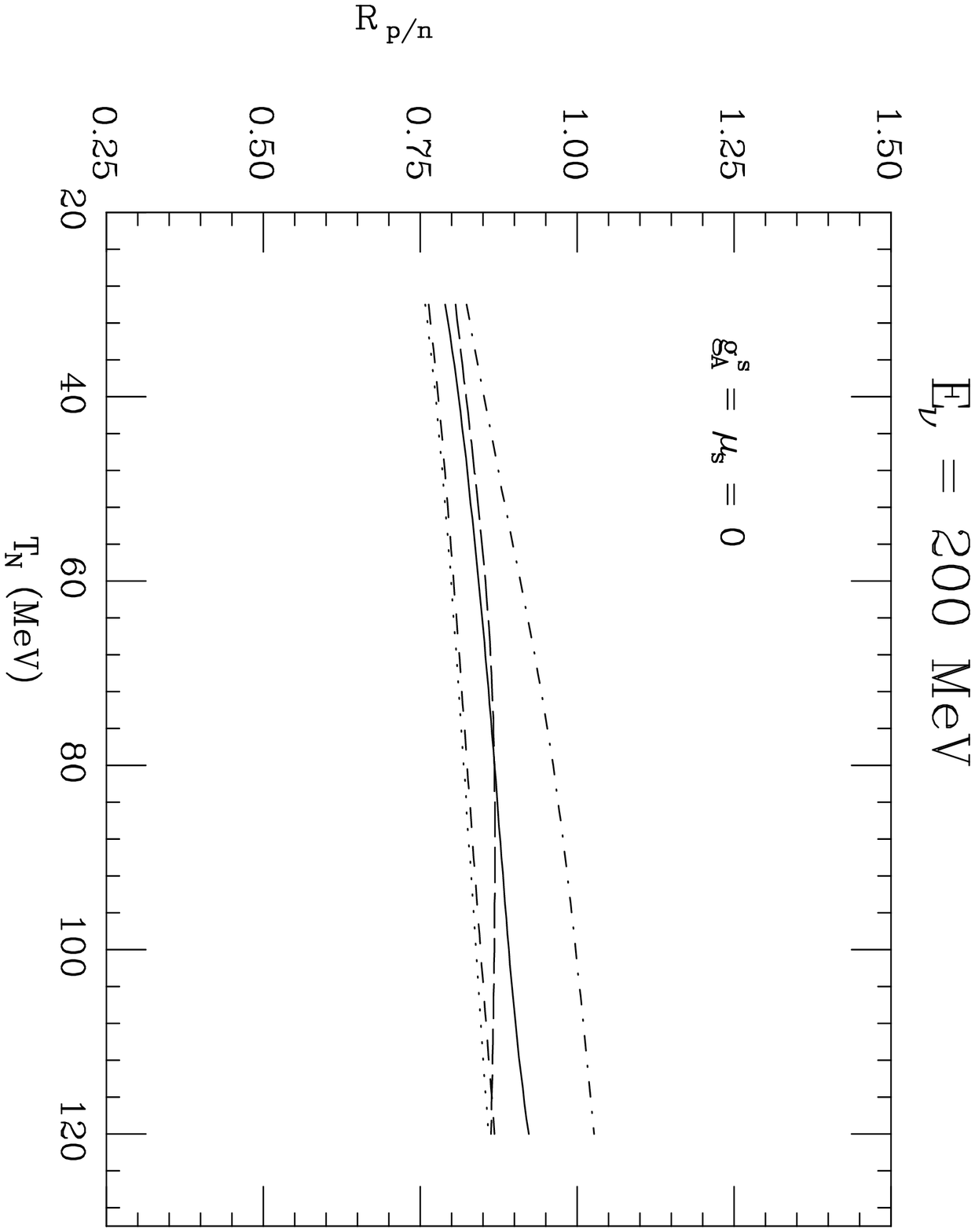,width=0.9\textwidth}}
\end{center}
\end{minipage}
\begin{center}
\Large Figure~\ref{fig06}
\end{center}

\begin{minipage}[p]{\textwidth}
\begin{center}
\mbox{\epsfig{file=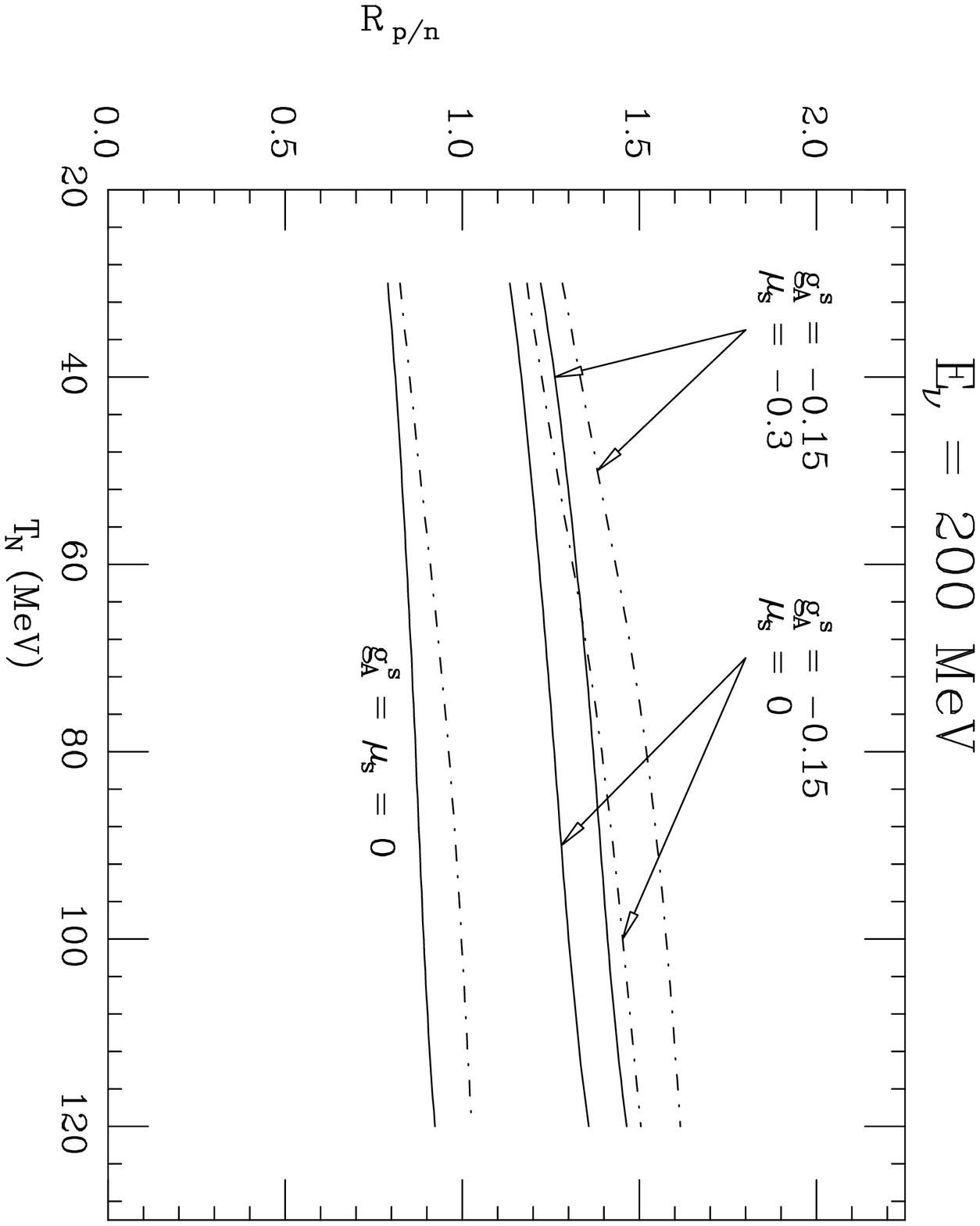,width=0.9\textwidth}}
\end{center}
\end{minipage}
\begin{center}
\Large Figure~\ref{fig07}
\end{center}

\begin{minipage}[p]{\textwidth}
\begin{center}
\mbox{\epsfig{file=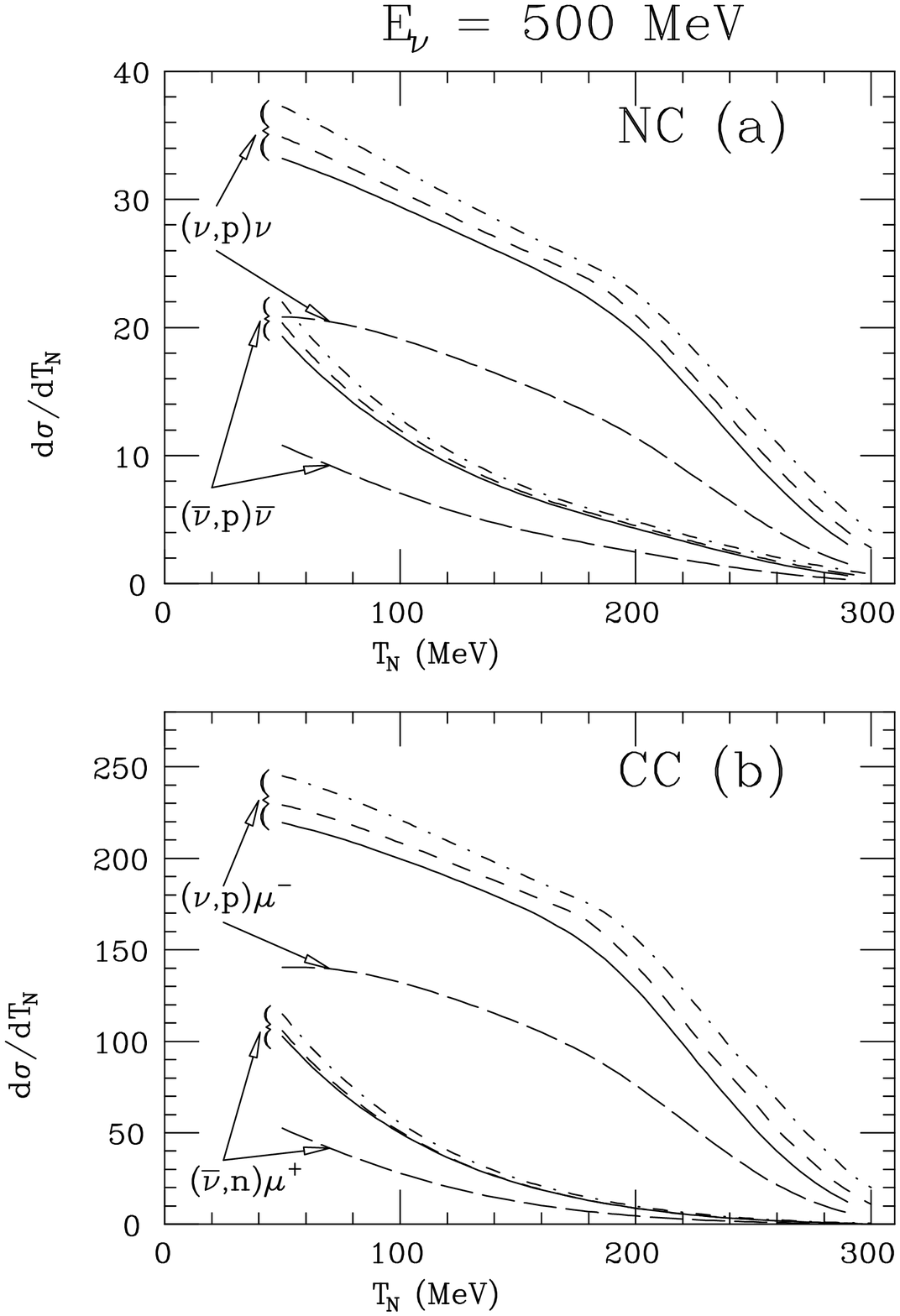,width=0.9\textwidth}}
\end{center}
\end{minipage}
\begin{center}
\Large Figure~\ref{fig08}
\end{center}

\begin{minipage}[p]{\textwidth}
\begin{center}
\mbox{\epsfig{file=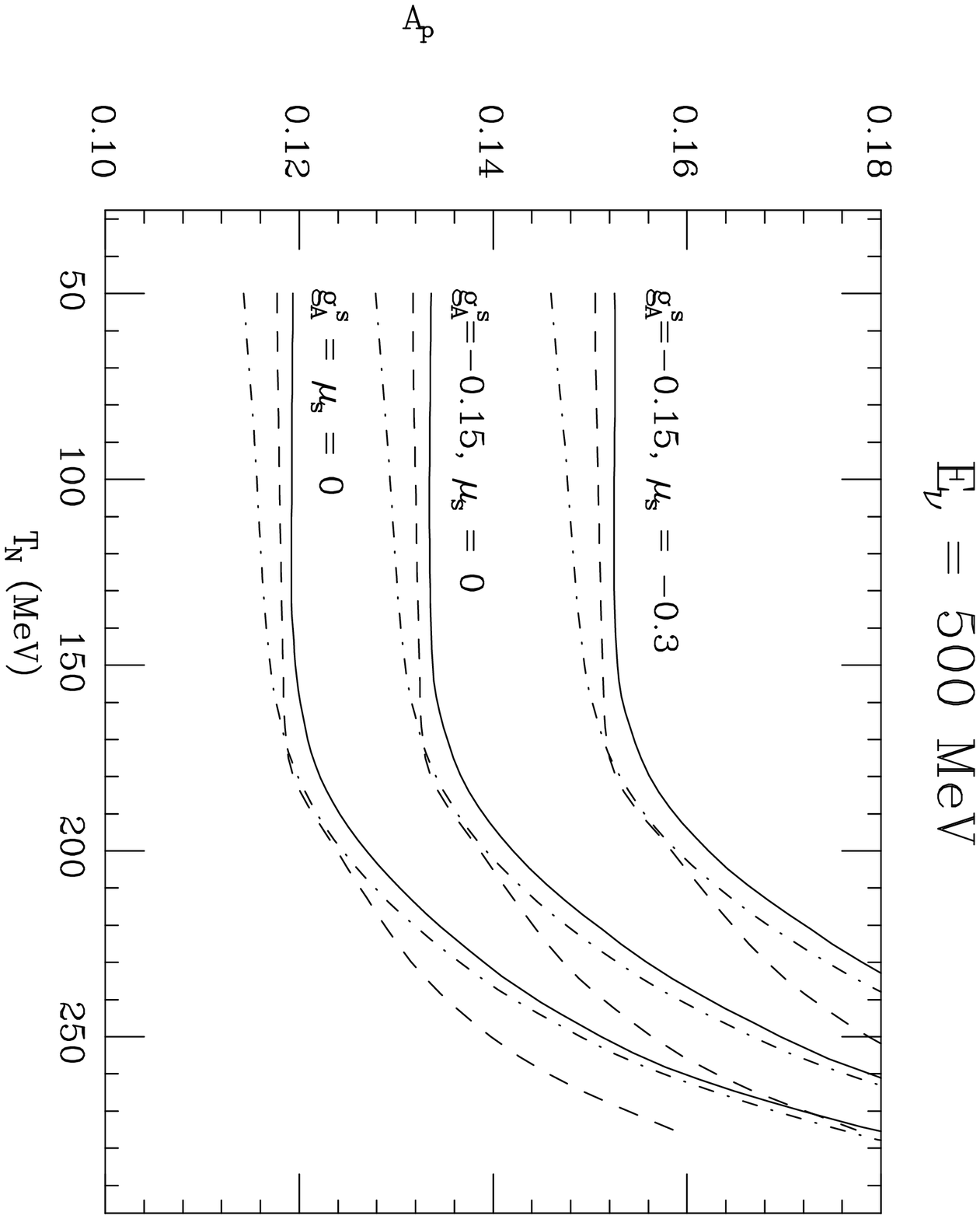,width=0.9\textwidth}}
\end{center}
\end{minipage}
\begin{center}
\Large Figure~\ref{fig09}
\end{center}

\begin{minipage}[p]{\textwidth}
\begin{center}
\mbox{\epsfig{file=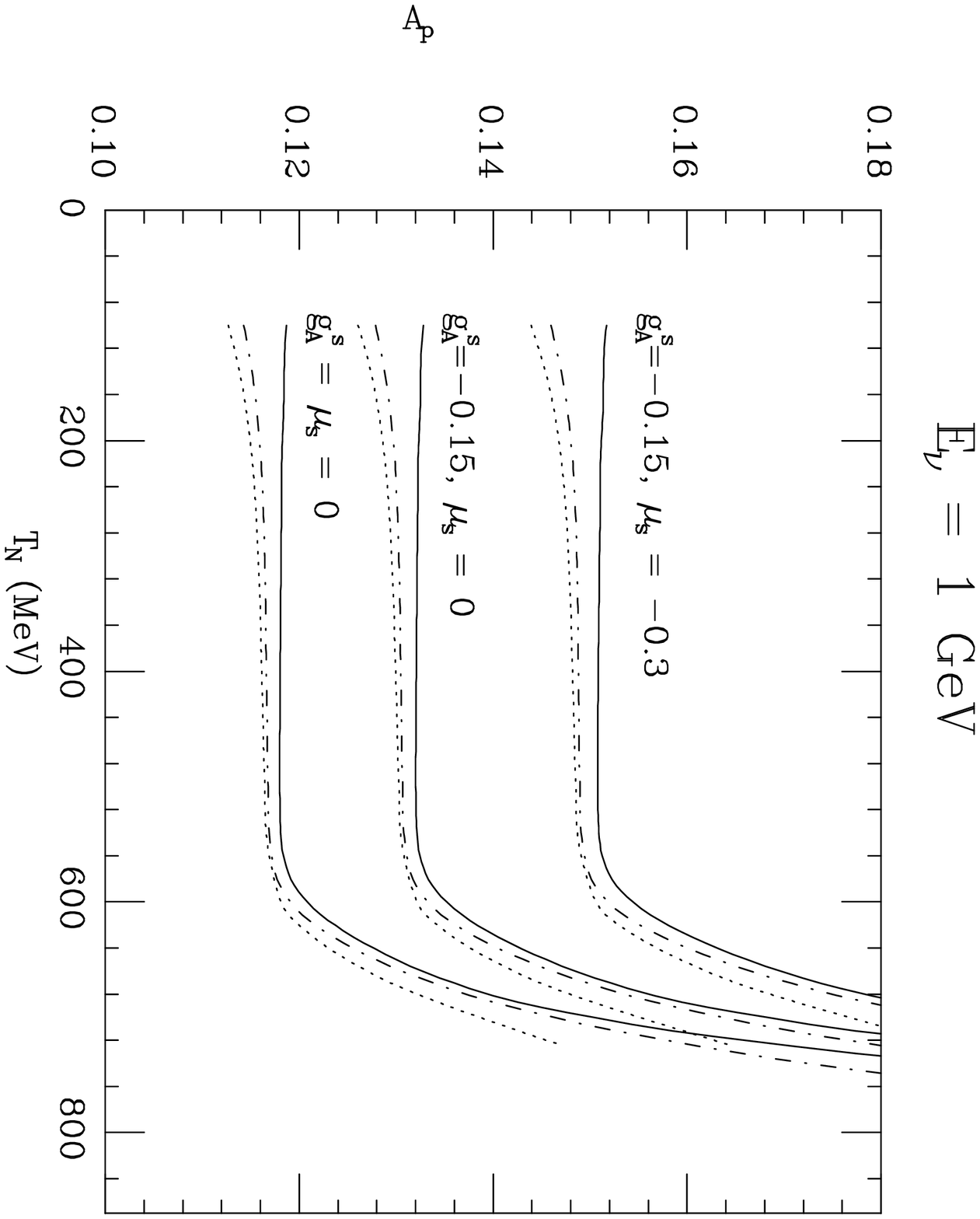,width=0.9\textwidth}}
\end{center}
\end{minipage}
\begin{center}
\Large Figure~\ref{fig10}
\end{center}

\begin{minipage}[p]{\textwidth}
\begin{center}
\mbox{\epsfig{file=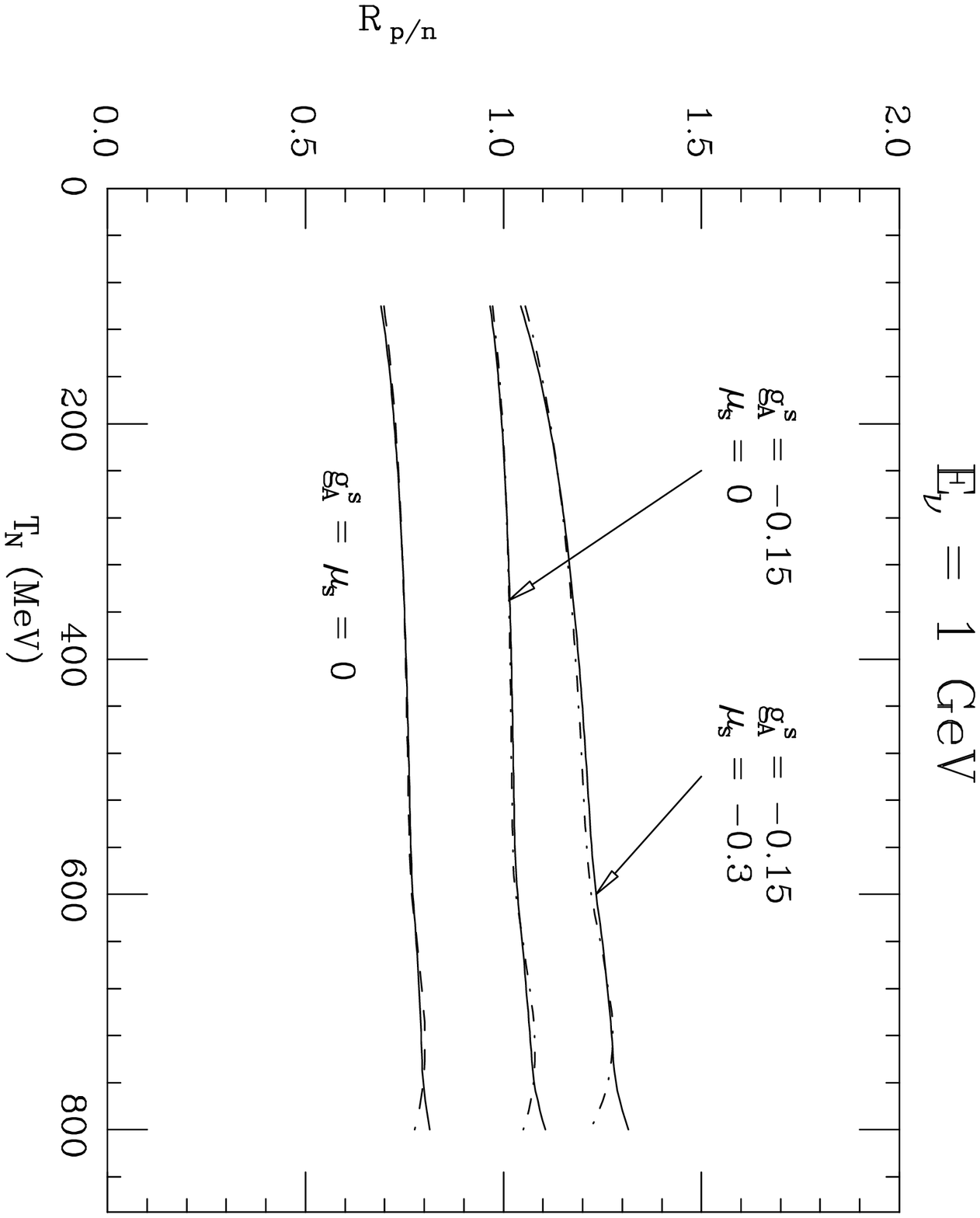,width=0.9\textwidth}}
\end{center}
\end{minipage}
\begin{center}
\Large Figure~\ref{fig11}
\end{center}

\begin{minipage}[p]{\textwidth}
\begin{center}
\mbox{\epsfig{file=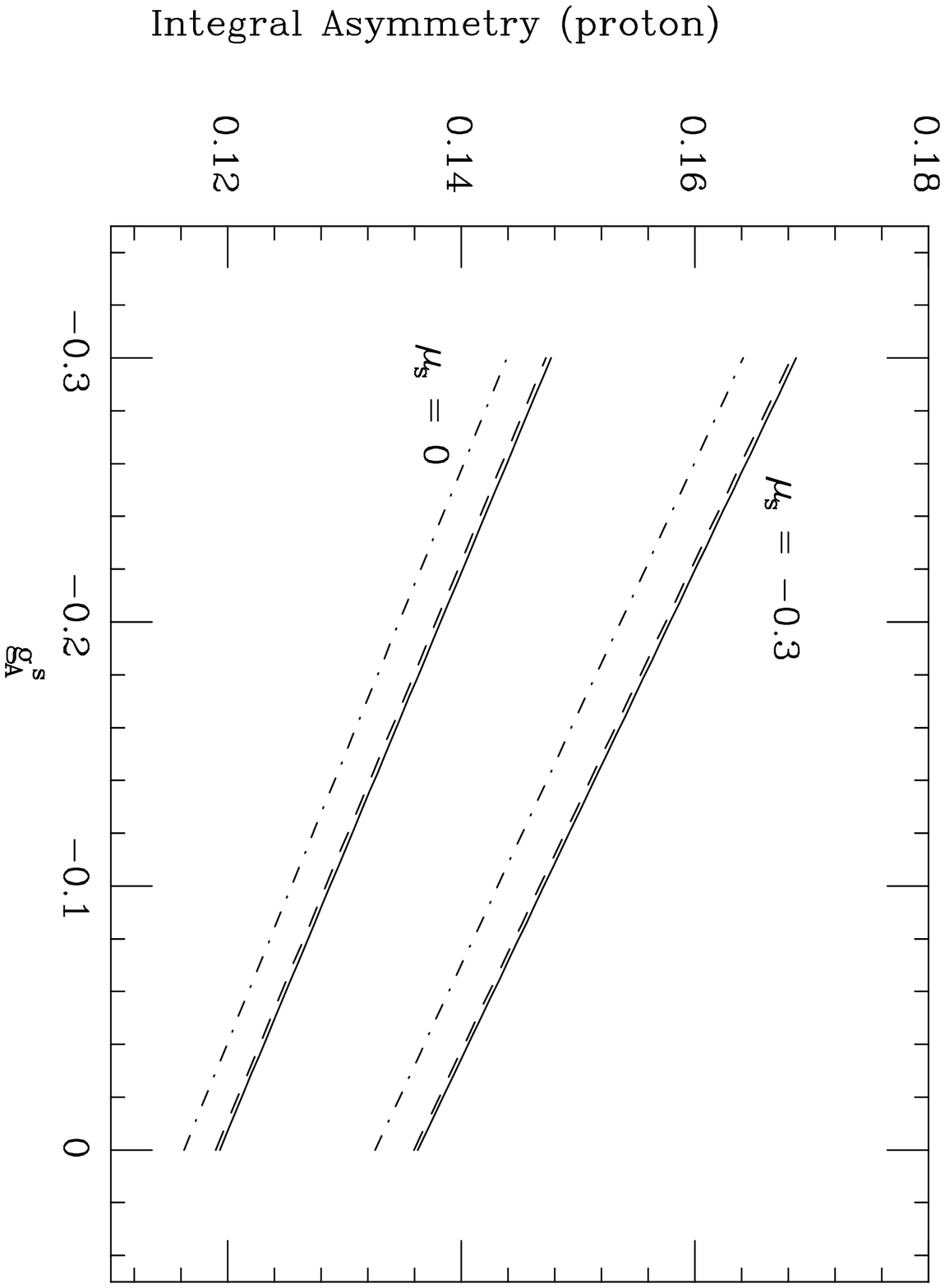,width=0.9\textwidth}}
\end{center}
\end{minipage}
\begin{center}
\Large Figure~\ref{fig12}
\end{center}

\end{document}